# Steepest-entropy-ascent nonequilibrium quantum thermodynamic framework to model chemical reaction rates at an atomistic level

**G. P. Beretta**[1] **and M. R. von Spakovsky**[2]

[1] Mechanical and Industrial Engineering Department
Università di Brescia, via Branze 38, 25123 Brescia, ITALY

[2] Center for Energy Systems Research, Mechanical Engineering Department
Virginia Polytechnic Institute and State University, Blacksburg, VA, U.S.A.

E-mail: gianpaolo.beretta@unibs.it and vonspako@vt.edu

**Abstract.** The steepest entropy ascent (SEA) dynamical principle provides a general framework for modeling the dynamics of nonequilibrium (NE) phenomena at any level of description, including the atomistic one. It has recently been shown to provide a precise implementation and meaning to the maximum entropy production principle and to encompass many well-established theories of nonequilibrium thermodynamics into a single unifying geometrical framework. Its original formulation in the framework of quantum thermodynamics (QT) assumes the simplest and most natural Fisher-Rao metric to geometrize from a dynamical standpoint the manifold of density operators, which represent the thermodynamic NE states of the system. This simplest SEAQT formulation is used here to develop a general mathematical framework for modeling the NE time evolution of the quantum state of a chemically reactive mixture at an atomistic level. The method is illustrated for a simple two-reaction kinetic scheme of the overall reaction $F + H_2 \Leftrightarrow HF + F$ in an isolated tank of fixed volume. However, the general formalism is developed for a reactive system subject to multiple reaction mechanisms. To explicitly implement the SEAQT nonlinear law of evolution for the density operator, both the energy and the particle number eigenvalue problems are set up and solved analytically under the dilute gas approximation. The system-level energy and particle number eigenvalues and eigenstates are used in the SEAQT equation of motion to determine the time evolution of the density operator, thus, effectively describing the overall kinetics of the reacting system as it relaxes towards stable chemical equilibrium. The predicted time evolution in the near-equilibrium limit is compared to the reaction rates given by a standard detailed kinetic model so as to extract the single time constant needed by the present SEA model.

## 1. Introduction

The principle of steepest entropy ascent (SEA) has recently been shown to provide a unifying framework for the construction of dynamical models of far-from-equilibrium phenomena at all levels of description, from the macroscopic to the microscopic [1-3]. As a consequence of the geometrization of the state space, which is a key requirement of the SEA construction, all models in the SEA family become intrinsically fundamental in the sense that they automatically enjoy a built-in strong form of compatibility with the second law of thermodynamics as well as with the Onsager reciprocity principle in the near equilibrium regime. In this paper, we demonstrate how the SEA principle can be embedded in an *ab initio* quantum chemical description of the kinetics of a simple set of chemical reactions and how it allows one to gain important insight into the reaction rates even in the far-from-equilibrium regime typical of chemical kinetics.

The SEA principle was first introduced as part of early attempts to develop a theory of quantum thermodynamics (QT) by complementing the postulates of quantum mechanics (QM) with the second law of thermodynamics [4-15]. The original idea behind the SEAQT equation of motion was to extend the linear Hamiltonian dynamics of zero-entropy QM, i.e., the Schrödinger equation, to a non-linear dynamics for the nonzero-entropy domain of QT in order to be able to describe systems undergoing nonequilibrium relaxation processes in which the entropy increases spontaneously due to internal dissipation. Whereas it originated from an extreme [16] and adventurous [17] physical ansatz about the meaning of entropy and irreversibility along the lines of thought that in the 1970's and 80's were also intensely sought after by Ilya Prigogine and coworkers [18], the strength of the geometrical basis of the SEA mathematical construction was immediately recognized to also provide a powerful tool for general nonequilibrium modeling, extending its usefulness beyond the original framework for which it was developed [19-23]. Although in recent years the field of quantum thermodynamics has developed in a number of different directions [24], the geometrical simplicity of the SEA idea has not lost its original appeal. Not only was it essentially rediscovered in QT [25, 26] several years after its original introduction [8] in an attempt to design a nonlinear quantum evolution with maximal entropy production identical with the original, but it has also been implicitly adopted [27] for the dissipative component of nonequilibrium evolution in the construction of the general equation for the nonequilibrium reversible-irreversible coupling known as GENERIC [28, 29]. The latter has been very successful in equipping models of, for example, complex fluids with a built-in thermodynamic consistency by emphasizing the interplay between the dissipative and nondissipative components of the dynamics and by applying the idea, also originated in the early 80's, of extending the bracket formalism to dissipative phenomena [30-33]. In more recent years, the mathematical community has somewhat independently adopted, developed, formalized, and generalized the geometrical idea of SEA to construct a variational framework for the mathematical theory of partial differential equations describing flows on metric-measure spaces currently referred to as "gradient flows" [34, 35].

The SEA mathematical framework is based on the observation that the description of dissipation in most well-established theories of nonequilibrium can be viewed essentially as particular implementations of the maximum entropy production principle (MEPP) [1]. The SEAQT equation of motion that we consider as the basis of our kinetic model is the original version developed for a quantum level description of dissipation. For pure (zero-entropy) density operators, the SEAQT equation of motion reduces to the standard Schrödinger equation of motion, which, of course, features no dissipation. However, for mixed density operators, the usual nondissipative Hamiltonian dynamics described by the standard (linear, unitary) von Neumann term in the evolution equation must instead compete with an additional (orthogonal) "pull" in the SEA direction, resulting from the non-standard (nonlinear, non-unitary) dissipative term which characterizes the SEAQT equation of motion. The resulting smooth, constant energy time evolution of the density operator determines the full NE path, which allows a determination of the time dependences of all the NE thermodynamic properties (e.g., composition, chemical potentials, affinities, reaction rates) including, of course, the NE entropy.

Since the system we model is isolated, the entropy can only increase in time and this irreversible increase emerges as the result of a spontaneous redistribution of the energy among the available energy eigenlevels until the final (maximum entropy) stable equilibrium distribution is reached, namely, the Gibbs-Boltzmann canonical distribution described by the maximum entropy density operator. The SEAQT equation of motion guarantees the (second-law, thermodynamic compatibility) requirement of non-negativity of the entropy generation along the entire smooth trajectory in state space for any initial density operator regardless of how far it starts from thermodynamic equilibrium. Moreover, since the dynamical equation in SEAQT implements the principle of MEPP, its application to chemical kinetics is conceptually consistent with the ideas put forward by Ziegler [36] concerning the thermodynamic consistency of the standard model of chemical kinetics.

In this paper, we construct a detailed, fully quantum, mathematical formalism for the application of the SEAQT dynamics to modeling chemical reaction rates. Our focus here is on chemically reactive systems at very small scales, i.e., on an isolated, chemically reactive mixture of very few molecules subject to $\tau$ active reaction mechanisms. The method we present has already been implemented by the research group at Virginia Tech to test its ability to model NE time evolutions in reactive systems [3, 37-40]. The *Appendix* of the present paper, contributed by Omar Al-Abassi, presents and discusses preliminary numerical results for a two-reaction model of the overall reaction $F + H_2 = HF + F$ that we adopt as a case-study throughout the paper to provide a concrete illustration of the formalism. The results shown in the Appendix are very encouraging and show that the method has the potential to provide new insights into far NE kinetics and to grow into a new tool for purely *ab-initio* approaches as well as for meso- and macroscopic modeling.

In modeling the NE time evolution of the state of these systems by means of the non-linear SEAQT equation of motion, both the system energy and the particle number eigenvalue problems must be solved. This establishes the so-called energy and particle number eigenstructure of the system, i.e., the landscape of quantum eigenstates available to the system. In this landscape, the SEAQT equation of motion determines the unique thermodynamic path taken by the density operator, which represents the thermodynamic state of the system at every instant of time, as it evolves from an arbitrary initial non-equilibrium state to the corresponding stable chemical equilibrium state (uniquely fixed by the initial state).

The paper is structured as follows. In *Section* 2, we define the quantum kinematics of the model by constructing the Hamiltonian and particle-occupation number operator for their formal eigenvalue problems in a way that implements the usual stoichiometric proportionality relations of the standard model of chemical kinetics. Using a two-reaction mechanism as illustration, we construct all the details of the Hilbert space of the system that are necessary to implement the model. In *Section* 3, we write the SEAQT equation of motion and work out its explicit form in terms of energy occupation probabilities under the simplifying assumption that the initial density operator commutes with the Hamiltonian. In *Section* 4 we present our conclusions and in the *Appendix*, contributed by Omar Al-Abassi, we present and discuss some numerical results that show the peculiarities of the proposed new method.

## 2. Quantum Thermodynamic Kinematics. Modeling Framework for a Reactive System

The thermodynamic system under consideration is an isolated, reacting mixture consisting of $r$ species $A_i$ contained in a tank, the walls of which form the boundaries of the system and isolate the mixture from its surroundings. Since the system is isolated, it experiences no heat, work, or mass interactions. The $\tau$ active reaction mechanisms are expressed as

$$\sum_{i=1}^{r} v_{il} A_i = 0 \quad l = 1,...,\tau \tag{1}$$

where $v_{il}$ is the stoichiometric or reaction coefficient for species $A_i$ in reaction mechanism "*l*".

Three systems of equations govern the evolution in time of the thermodynamic state of this system. The first two, discussed in the present section, are the energy and particle number eigenvalue problems, which are used to establish the time-independent (kinematic) eigenstructure of the system defining the mathematical framework for the equation of motion (e.g., that of SEAQT) that forms a third system of governing (dynamical) equations discussed in *Section* 3.

*2.1 Energy and Particle Occupation Number Eigenvalue Problems*

The energy eigenvalue problem, which must be solved for this system once the system-level Hamiltonian operator *H* is defined, is as follows:

$$H|\xi_{sq_s}\rangle = E_{sq_s}|\xi_{sq_s}\rangle \quad s=1,...,C \quad q_s=1,...,L_s \tag{2}$$

where $C$ is the number of subspaces of compatible compositions and $L_s$ the dimension of subspace $s$. In Eq. (2), $E_{sq_s}$ and $|\xi_{sq_s}\rangle$ are the system-level energy eigenvalues and eigenvectors, respectively. The dimension of the overall Hilbert space $\mathscr{H}$ of the system is $L = \sum_{s=1}^{C} L_s$. A Hilbert as opposed to Fock space is assumed since the framework presented is based on the assumption that, consistent with the earlier assumption of an isolated system, the number of atoms is fixed (i.e., is conserved) and always known. However, as we will see later, the structure of the assumed Hilbert space reflects the fact that the species particles (molecules) are in most states fluctuating in number because of the presence of the chemical reaction mechanism(s). In other words, the composition in terms of different kinds of molecules must be described in terms of probabilities associated with their possible numbers, while the composition in terms of different kinds of atoms that are assembled and disassembled by the reaction mechanism(s) to form different molecules is fixed.

Connected with Eq. (2), a second set of governing equations is given by the particle occupation number eigenvalue problems

$$N_{A_i j_i}|\xi_{sq_s}\rangle = \alpha_{ij_i}^{sq_s}|\xi_{sq_s}\rangle \quad s=1,...,C \quad q_s=1,...,L_s \quad i=1,...,r \quad j_i=1,...,M_i \tag{3}$$

where $r$ is the number of species, $N_{A_i j_i}$ the $A_i$-particles-in-the-$j_i^{th}$-internal-level occupation number operator, and $\alpha_{ij_i}^{sq_s}$ the $A_i$-particles-in-the-$j_i^{th}$-internal-level eigenvalue for the $q_s^{th}$ combination in the $s^{th}$ compatible composition. $M_i$ is the number of eigenvectors of the one-$A_i$-particle internal Hamiltonian operator. As is shown below, the $\alpha_{ij_i}^{sq_s}$ are related to the Bose-Einstein or Fermi-Dirac occupation numbers defined below. The relationship between the system-level energy eigenvectors $|\xi_{sq_s}\rangle$ and the one-particle energy eigenvectors of the different species is developed below as well. As a first step and to clarify the meaning of Eqs. (2) and (3), we examine the structure of the Hilbert space they imply as required for our chemically reactive system.

*2.2 Compatible Compositions and the Structure of the Hilbert Space*

To define the Hilbert space and the set of eigenvectors that span that space, the initial amount $n_{ia}$ for each species $A_i$ in the reacting mixture is first related to the set of compatible amounts $n_i(\boldsymbol{\varepsilon}_s)$ via the proportionality relations [41]

$$n_{is} = n_i(\boldsymbol{\varepsilon}_s) = n_{ia} + \sum_{l=1}^{\tau} v_{il}\varepsilon_{ls} \geq 0 \tag{4}$$

$$= n_{ia} + \boldsymbol{v}_i \cdot \boldsymbol{\varepsilon}_s \tag{5}$$

where the $n_i$ are eigenvalues of the $A_i$-particles number operator (defined in Eq. (22) below), $\varepsilon_{ls}$ is the eigenvalue of the reaction coordinate operator (defined in Eq. (21) below) for reaction "$l$" corresponding to the $s^{th}$ compatible composition, $\boldsymbol{v}_i$ the set of stoichiometric coefficients for species $A_i$ in each of the $\tau$ reaction mechanisms, and $\boldsymbol{\varepsilon}_s$ the set of reaction coordinate eigenvalues identifying the $s^{th}$ compatible composition given by

$$\boldsymbol{\varepsilon}_s = (\varepsilon_{1s},...,\varepsilon_{ls},...,\varepsilon_{\tau s}) \tag{6}$$

The values of the reaction coordinates $\boldsymbol{\varepsilon}_s$ cannot be assigned arbitrarily because through the proportionality relations (Eq. (4)), they determine the species amounts eigenvalues

$$\boldsymbol{n}_s = (n_{1s},...,n_{is},...,n_{rs}) \tag{7}$$

which by definition must be non-negative integers. Therefore, the combinations of reaction coordinates that are compatible with the given initial amounts $\boldsymbol{n}_a = (n_{1a},...,n_{ia},...,n_{ra})$ are finite in number. We denote this number of compatible compositions by $C$ and the set of all compatible combinations of reaction coordinates by

$$\{\boldsymbol{\varepsilon}_s\} = \{\varepsilon_{ls} | \ l=1,...,\tau \quad s=1,...,C\} \tag{8}$$

and the corresponding set of compositions by

$$\{\mathbf{n}_s\} = \{n_{is} \mid i = 1,...,r \quad s = 1,...,C\} \tag{9}$$

Note that the $n_{is}$ are related to the $\alpha_{ij_i}^{sq_s}$ eigenvalues appearing in our particle number eigenvalue problems, Eq. (3), by

$$n_{is} = \sum_{j_i}^{M_i} \alpha_{ij_i}^{sq_s} \quad \text{for every } q_s = 1,...,L_s \tag{10}$$

where

$$0 \leq \alpha_{ij_i}^{sq_s} \leq n_{is} \quad \text{for every } q_s = 1,...,L_s \tag{11}$$

If we assume that identical particles are distinguishable, the overall system Hilbert space, H, is now defined in terms of the one-particle spaces $\mathcal{H}^{A_i}$ of the different species $A_i$ as follows:

$$\mathcal{H} = \bigoplus_{s=1}^{C} \bigotimes_{i=1}^{r} \left(\mathcal{H}^{A_i}\right)^{\otimes n_{is}} = \bigoplus_{s=1}^{C} \mathcal{H}_s \tag{12}$$

where $\mathcal{H}_s$ is the Hilbert space that we would use to describe a closed system with fixed composition equal to $\mathbf{n}_s$.

As an illustration of how such a space is constructed, we consider a reacting mixture system with the following two-reaction mechanisms:

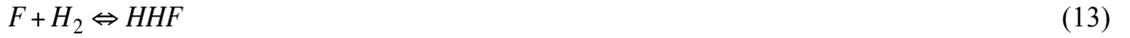
$$F + H_2 \Leftrightarrow HHF \tag{13}$$

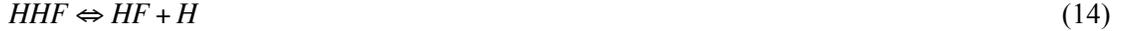
$$HHF \Leftrightarrow HF + H \tag{14}$$

As discussed in the *Appendix*, the *HHF* in this mechanism plays the role of the activated complex. Therefore, these two-reaction mechanisms are taken here as the simplest way to exemplify the standard model of chemical kinetics and to justify the multi-reaction framework that we develop in this paper in order to implement the SEAQT model.

The $C$ sets of compatible values of the $\tau = 2$ parameters $\boldsymbol{\varepsilon}_s = \{\varepsilon_{1s}, \varepsilon_{2s}\}$ are found by imposing the non-negativity condition on the values of $n_{is}$ obtained from Eq. (4) for given initial amounts. The procedure is illustrated in Table 1 where the initial amounts chosen for $F$, $H_2$, $HHF$, $HF$, and $H$ are 4, 1, 0, 0, and 0, respectively. In this table, the final column represents the bounds on $\varepsilon_1$ and $\varepsilon_2$ imposed by the non-negativity constraint on $n_{is}$. A plot of these bounds is given in Figure 1 from which one can conclude that there are only $C = 3$ combinations of values compatible with the integer and non-negativity conditions, namely,

$$\{\boldsymbol{\varepsilon}_s\} = \{\boldsymbol{\varepsilon}_1 = (0,0), \boldsymbol{\varepsilon}_2 = (1,0), \boldsymbol{\varepsilon}_3 = (1,1)\} \tag{15}$$

and the corresponding set of compatible compositions is

$$\{\mathbf{n}_s\} = \{\mathbf{n}_1 = (4,1,0,0,0), \mathbf{n}_2 = (3,0,1,0,0), \mathbf{n}_3 = (3,0,0,1,1)\} \tag{16}$$

The relationship between $\mathbf{n}_s$ and $\boldsymbol{\varepsilon}_s$ (i.e., between Eqs. (15) and (16)) is illustrated in Table 2. With these values, the Hilbert space for this system is written as

$$\mathcal{H} = \left(\mathcal{H}^{A_1}\right)^{\otimes 4} \otimes \left(\mathcal{H}^{A_2}\right) \oplus \left(\mathcal{H}^{A_1}\right)^{\otimes 3} \otimes \left(\mathcal{H}^{A_3}\right) \oplus \left(\mathcal{H}^{A_1}\right)^{\otimes 3} \otimes \left(\mathcal{H}^{A_4}\right) \otimes \left(\mathcal{H}^{A_5}\right) \tag{17}$$

$$= \mathcal{H}_1 \oplus \mathcal{H}_2 \oplus \mathcal{H}_3 \tag{18}$$

In general,

$$\mathcal{H}_s = \bigotimes_{i=1}^{r} \left(\mathcal{H}^{A_i}\right)^{\otimes n_{is}} \tag{19}$$

is the subspace associated with the $s^{th}$ compatible composition. The projection operators $P_{\mathcal{H}_s}$ determine a spectral resolution of the identity operator $I$ on the overall Hilbert space where

$$I = \sum_{s=1}^{C} P_{\mathcal{H}_s} \tag{20}$$

**Table 1.** Limits on the reaction coordinates for the two-reaction-mechanisms system imposed by the integer and non-negativity conditions on the eigenvalues of the species number operators.

| $i$ | $A_i$ | $\nu_{i1}$ | $\nu_{i2}$ | $n_{ia}$ | $n_{is}$ | $n_{is} \geq 0$ |
|---|---|---|---|---|---|---|
| 1 | F | -1 | 0 | 4 | $4 - \varepsilon_{1s}$ | $\varepsilon_{1s} \leq 4$ |
| 2 | $H_2$ | -1 | 0 | 1 | $1 - \varepsilon_{1s}$ | $\varepsilon_{1s} \leq 1$ |
| 3 | HHF | 1 | -1 | 0 | $\varepsilon_{1s} - \varepsilon_{2s}$ | $\varepsilon_{1s} \geq \varepsilon_{2s}$ |
| 4 | HF | 0 | 1 | 0 | $\varepsilon_{2s}$ | $\varepsilon_{2s} \geq 0$ |
| 5 | H | 0 | 1 | 0 | $\varepsilon_{2s}$ | $\varepsilon_{2s} \geq 0$ |

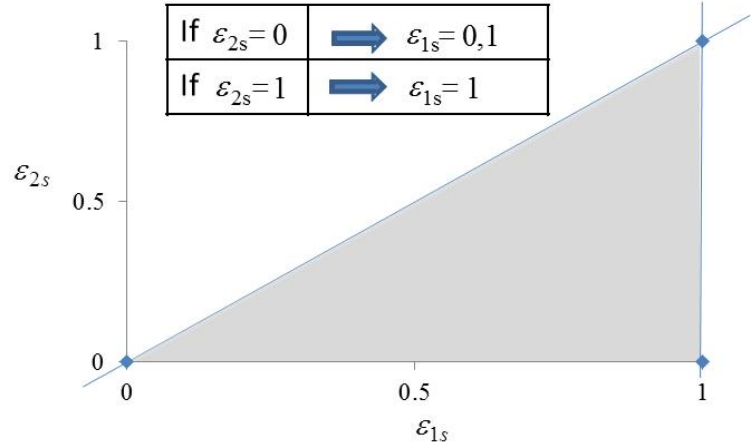

**Figure 1.** Shaded area graphically defines the limits defined in Table 1.

**Table 2.** The relationship between the $n_s$ and the $\varepsilon_s$ for the two-mechanisms example.

| | $s$ | 1 | 2 | 3 |
|---|---|---|---|---|
| $l$ | $\varepsilon_{ls}$ | $\varepsilon_{l1}$ | $\varepsilon_{l2}$ | $\varepsilon_{l3}$ |
| 1 | $\varepsilon_{1s}$ | 0 | 1 | 1 |
| 2 | $\varepsilon_{2s}$ | 0 | 0 | 1 |
| $i$ | $n_{is}$ | $n_{i1}$ | $n_{i2}$ | $n_{i3}$ |
| 1 | $n_{1s}$ | 4 | 3 | 3 |
| 2 | $n_{2s}$ | 1 | 0 | 0 |
| 3 | $n_{3s}$ | 0 | 1 | 0 |
| 4 | $n_{4s}$ | 0 | 0 | 1 |
| 5 | $n_{5s}$ | 0 | 0 | 1 |

and the expectation value $\langle P_{\mathcal{H}_s} \rangle = Tr(\rho P_{\mathcal{H}_s})$ represents the probability that the system is found in the $s^{th}$ compatible composition. Here $\rho$ is the "state" operator, which represents the state of the system. Of course, the operator $\rho$ is a linear, self-adjoint, non-negative definite, unit-trace operator on $\mathcal{H}$ (i.e., a linear operator with real, non-negative eigenvalues that sum up to unity). At a given instant of time, it is a complete representation of the thermodynamic non-equilibrium or equilibrium state of the system in the sense that it determines the expectation values of all observables, including, in particular, how the energy and the particles of the system are distributed amongst the various system energy and particle number eigenlevels.

*2.3 Number of Particles and Reaction Coordinate Operators*

The projection operators $P_{\mathcal{H}_s}$ determine the set $\boldsymbol{\mathcal{E}}$ of reaction coordinate operators

$$\boldsymbol{\mathcal{E}} = \sum_{s=1}^{C} \boldsymbol{\varepsilon}_s P_{\mathcal{H}_s} \quad \text{i.e.,} \quad \mathcal{E}_l = \sum_{s=1}^{C} \varepsilon_{ls} P_{\mathcal{H}_s} \quad l = 1,...,\tau \tag{21}$$

and the $A_i$-particles number operator $N_{A_i}$

$$N_{A_i} = \sum_{s=1}^{C} n_{is} P_{\mathcal{H}_s} = n_{ia} I + \boldsymbol{v}_i \cdot \boldsymbol{\mathcal{E}} \tag{22}$$

Moreover, the relationship between $N_{A_i}$ and the $A_i$-particles-in-the-$j_i^{th}$-internal-level occupation number operator $N_{A_i j_i}$ is given by

$$N_{A_i} = \sum_{j_i=1}^{M_i} N_{A_i j_i} \tag{23}$$

where $j_i = 1,...,M_i$ labels the eigenvectors of the one-$A_i$-particle internal Hamiltonian operator on $\mathcal{H}^{A_i}$.

For the two-reaction-mechanisms example, the set of reaction coordinate operators is

$$\boldsymbol{\mathcal{E}} = \{\mathcal{E}_1 = (0)P_{\mathcal{H}_1} + (1)P_{\mathcal{H}_2} + (1)P_{\mathcal{H}_3}, \mathcal{E}_2 = (0)P_{\mathcal{H}_1} + (0)P_{\mathcal{H}_2} + (1)P_{\mathcal{H}_3}\} \tag{24}$$

$$= \{\mathcal{E}_1 = P_{\mathcal{H}_2} + P_{\mathcal{H}_3}, \mathcal{E}_2 = P_{\mathcal{H}_3}\} \tag{25}$$

while the $A_i$-particles number operators are written as

$$N_F = 4I - \mathcal{E}_1 = 4I - P_{\mathcal{H}_2} - P_{\mathcal{H}_3} \tag{26}$$

$$N_{H_2} = I - \mathcal{E}_1 = I - P_{\mathcal{H}_2} - P_{\mathcal{H}_3} \tag{27}$$

$$N_{HHF} = \mathcal{E}_1 - \mathcal{E}_2 = P_{\mathcal{H}_2} \tag{28}$$

$$N_{HF} = \mathcal{E}_2 = P_{\mathcal{H}_3} \tag{29}$$

$$N_H = \mathcal{E}_2 = P_{\mathcal{H}_3} \tag{30}$$

where we have used Eq. (15) and the fact that

$$\boldsymbol{v}_1 = (-1,0), \quad \boldsymbol{v}_2 = (-1,0), \quad \boldsymbol{v}_3 = (1,-1), \quad \boldsymbol{v}_4 = (0,1), \quad \boldsymbol{v}_5 = (0,1) \tag{31}$$

Now, returning to the general formulation, the expectation values for the number of particles operators and those for the reaction coordinates operators are given by

$$\langle N_{A_i} \rangle = \sum_{s=1}^{C} n_{is} \text{Tr}(\rho P_{\mathcal{H}_s}) \quad \text{or} \quad \langle \boldsymbol{N} \rangle = \sum_{s=1}^{C} \boldsymbol{n}_s \text{Tr}(\rho P_{\mathcal{H}_s}) \tag{32}$$

and $\quad \langle \mathcal{E}_{ls} \rangle = \sum_{s=1}^{C} \varepsilon_{ls} \text{Tr}(\rho P_{\mathcal{H}_s}) \quad \text{or} \quad \langle \boldsymbol{\mathcal{E}} \rangle = \sum_{s=1}^{C} \boldsymbol{\varepsilon}_s \text{Tr}(\rho P_{\mathcal{H}_s}) \tag{33}$

As seen in the previous section, each trace term in these expressions represents the probability that the system is found in the $s^{th}$ compatible composition. For simplicity of notation, we denote these probabilities by

$$w_s = \text{Tr}(\rho P_{\mathcal{H}_s}) \quad \text{and} \quad \sum_{s=1}^{C} w_s = 1 \tag{34}$$

so that the expectation values $\langle \boldsymbol{N} \rangle$ and $\langle \boldsymbol{\mathcal{E}} \rangle$ are rewritten as

$$\langle N \rangle = \sum_{s=1}^{C} \boldsymbol{n}_s w_s \tag{35}$$

and  $$\langle \boldsymbol{\mathcal{E}} \rangle = \sum_{s=1}^{C} \boldsymbol{\varepsilon}_s w_s \tag{36}$$

For the two-reaction-mechanisms example,
$$\langle \mathcal{E}_1 \rangle = w_2 + w_3 \tag{37}$$
$$\langle \mathcal{E}_2 \rangle = w_3 \tag{38}$$
and  $$\langle N_F \rangle = 4w_1 + 3w_2 + 3w_3 \tag{39}$$
$$\langle N_{H_2} \rangle = w_1 \tag{40}$$
$$\langle N_{FHH} \rangle = w_2 \tag{41}$$
$$\langle N_{FH} \rangle = w_3 \tag{42}$$
$$\langle N_H \rangle = w_3 \tag{43}$$

*2.4 Hamiltonian Operator*

In order to relate the Hamiltonian operator $H$ for the overall system to the one-particle Hamiltonians and the interaction Hamiltonians, we make some standard assumptions. Let us start with the Hamiltonians $H_s$ for the fixed composition space $\mathcal{H}_s$. It is convenient to factor the one particle space $\mathcal{H}^{A_i}$ into its translational, rotational, vibrational, and electronic parts (additional divisions are possible) so that

$$\mathcal{H}^{A_i} = \mathcal{H}_{tr}^{A_i} \otimes \mathcal{H}_{rot}^{A_i} \otimes \mathcal{H}_{vib}^{A_i} \otimes \mathcal{H}_{el}^{A_i} = \mathcal{H}_{tr}^{A_i} \otimes \mathcal{H}_{int}^{A_i} \tag{44}$$

Here for simplicity we group the non-translational factor spaces into a single one that we name "internal". Thus, we may write

$$\mathcal{H}_s = \bigotimes_{i=1}^{r} \left( \mathcal{H}^{A_i} \right)^{\otimes n_{is}} = \bigotimes_{i=1}^{r} \left( \mathcal{H}_{tr}^{A_i} \right)^{\otimes n_{is}} \bigotimes_{i=1}^{r} \left( \mathcal{H}_{int}^{A_i} \right)^{\otimes n_{is}} = \mathcal{H}_s^{tr} \otimes \mathcal{H}_s^{int} \tag{45}$$

and for the Hamiltonian

$$H_s = H_s^{tr} \otimes I_s^{int} + I_s^{tr} \otimes H_s^{int} + V_s^{tr-int} \tag{46}$$

where $V_s^{tr-int}$ is the interaction term between the translational and internal degrees of freedom.

Next, we write the $A_i$-one-particle internal Hamiltonians $H_{int}^{A_i}$ associated with the $A_i$-particles-in-the-$j_i^{th}$-internal level occupation number operator $N_{A_i,j_i}$ as follows:

$$H_{int}^{A_i} \left| \varepsilon_{j_i}^{A_i} \right\rangle = e_{j_i}^{A_i} \left| \varepsilon_{j_i}^{A_i} \right\rangle \quad i = 1,\ldots,r \quad j_i = 1,\ldots,M_i \tag{47}$$

where $M_i$ represents the dimension of $\mathcal{H}_{int}^{A_i}$ and $e_{j_i}^{A_i}$ is the $A_i$-one-particle internal energy eigenvalue belonging to the $j_i^{th}$ internal energy eigenvector $\left| \varepsilon_{j_i}^{A_i} \right\rangle$. Since we are interested in modeling Bosons, we assume that the eigenvectors of the internal Hamiltonian $H_s^{int}$ for the composite are given by the separable combinations of the single particle eigenvectors that are invariant upon exchange of two identical particles. As a result, the order in the factorization is unimportant and all that counts are the occupation numbers $\alpha_{ij_i}^{sq_s^{int}}$ where $q_s^{int}$ labels the possible combinations. Thus,

$$\left| \xi_{sq_s^{int}}^{int} \right\rangle = \bigotimes_{i=1}^{r} \left( \bigotimes_{j_i=1}^{M_i} \left| \varepsilon_{j_i}^{A_i} \right\rangle^{\otimes \alpha_{ij_i}^{sq_s^{int}}} \right) \tag{48}$$

and its eigenvalues

$$E_{sq_s^{int}}^{int} = \sum_{i=1}^{r} \sum_{j_i=1}^{M_i} e_{j_i}^{A_i} \alpha_{ij_i}^{sq_s^{int}} \tag{49}$$

We develop this further in *Section* 2.6 below, showing how the eigenvectors of $H_s^{int}$ would be constructed using our example of the two-reaction-mechanism system.

Now in contrast to $H_s^{int}$, the translational Hamiltonian $H_s^{tr}$ is in general non-separable because of the intermolecular forces, which could be model, for example, using a Lennard-Jones pairwise potential. However, we can perform the standard change of variables that allows one to separate the "center of mass" of the group of $n_s$ particles from the "relative particles" with reduced mass. As a result, the interaction term depends only on the relative (or reduced) coordinates [42]. If we further assume, as in standard practice, that such an interaction term separates into a sum of terms each depending only on one of the relative coordinates, then the overall translational Hamiltonian separates into $n_s = \sum_{i=1}^{r} n_{is}$ terms. This procedure corresponds to performing a unitary transformation $T$ on the translational Hamiltonian operator $H_s^{tr}$ to obtain the operator

$$\hat{H}_s^{tr} = T H_s^{tr} T^{-1} = \sum_{j_s=1}^{n_s} \hat{H}_{j_s}^{s} \otimes \hat{I}_{\bar{j}_s} \tag{50}$$

and a new factorization of the Hilbert space $\hat{\mathscr{H}}_s^{tr}$

$$\hat{\mathscr{H}}_s^{tr} = \bigotimes_{j_s=1}^{n_s} \hat{\mathscr{H}}_{j_s} \tag{51}$$

such that for every $j_s$, $\hat{H}_{j_s}^{s}$ is an operator on $\hat{\mathscr{H}}_{j_s}$ and $\hat{I}_{\bar{j}_s}$ denotes the identity on the complementary space $\bigotimes_{j_p=1, j_p \neq j_s}^{n_s} \hat{\mathscr{H}}_{j_p}$.

The advantage of the above transformation is that the eigenvalue problem for $H_s^{tr}$ is decoupled into $n_s$ smaller eigenvalue problems

$$\hat{H}_{j_s}^{s} \left| \hat{\varepsilon}_{k_{sj_s}}^{sj_s} \right\rangle = e_{k_{sj_s}}^{sj_s} \left| \hat{\varepsilon}_{k_{sj_s}}^{sj_s} \right\rangle \quad s=1,...,C \quad j_s=1,...,n_s \quad k_{sj_s}=1,...,K_{sj_s} \tag{52}$$

and as a result,

$$\hat{H}_{j_s}^{s} = \sum_{k_{sj_s}=1}^{K_{sj_s}} e_{k_{sj_s}}^{sj_s} \left| \hat{\varepsilon}_{k_{sj_s}}^{sj_s} \right\rangle \left\langle \hat{\varepsilon}_{k_{sj_s}}^{sj_s} \right| \quad s=1,...,C \quad j_s=1,...,n_s \tag{53}$$

Substituting back into Eq. (50), we get $H_s^{tr}$, which can then be transformed back into the original variables to yield

$$H_s^{tr} = T^{-1} \hat{H}_s^{tr} T \tag{54}$$

The eigenvectors of this transformed Hamiltonian are, therefore, given by

$$\left| \xi_{sq_s^{tr}}^{tr} \right\rangle = T^{-1} \left( \bigotimes_{j_s=1}^{n_s} \left| \hat{\varepsilon}_{k_{sj_s}}^{sj_s} \right\rangle \right) \tag{55}$$

where we introduce the notation $q_s^{tr} = (k_{s1},...,k_{sj_s},...,k_{sn_s})$ and each $k_{sj_s} = 1,...,K_{sj_s}$. The corresponding energy eigenvalues are then

$$E_{sq_s^{tr}}^{tr} = e_{k_{s1}}^{s1} + ... + e_{k_{sj_s}}^{sj_s} + ... + e_{k_{sn_s}}^{sn_s} \tag{56}$$

It is noteworthy that the translational problem is infinite dimensional and, therefore, the values $K_{sj_s}$ correspond to some practical truncation made to render the problem numerically tractable. In addition, although the operators $H_s^{tr}$ and $\hat{H}_s^{tr}$ have the same set of eigenvalues, the eigenvectors of $\hat{H}_s^{tr}$ are factored, while those of $H_s^{tr}$ are in general not due to the effect of the unitary transformation $T^{-1}$ (see Eq. (50)). As a result, the set of quantum numbers $q_s^{tr}$ identifies a combination of independent modes in the center of mass and relative coordinate framework but a "collective mode" shared by all the particles when viewed from the untransformed set of variables corresponding to the original structure

$$\mathcal{H}_s^{tr} = \bigotimes_{i=1}^{r} \left( \mathcal{H}_{tr}^{A_i} \right)^{\otimes n_{is}} \tag{57}$$

Now, returning to the eigenvalue problem of the overall Hamiltonian $H_s$, if we neglect the $V_s^{tr-int}$ interaction, $H_s$ separates into the internal plus the translational problems we just addressed. Therefore, the eigenvectors associated with $H_s$ are

$$\left| \xi_{sq_s} \right\rangle = \left| \xi_{sq_s^{tr}}^{tr} \right\rangle \otimes \left| \xi_{sq_s^{int}}^{int} \right\rangle \tag{58}$$

and the eigenvalues

$$E_{sq_s} = E_{sq_s^{tr}}^{tr} + E_{sq_s^{int}}^{int} \tag{59}$$

The eigenvectors $\left| \xi_{sq_s} \right\rangle$ form a basis set for subspace $\mathcal{H}_s$. Denoting their 1D linear span by $\mathcal{H}_{sq_s}$, the subspace $\mathcal{H}_s$ is, therefore, written as

$$\mathcal{H}_s = \bigoplus_{q_s=1}^{L_s} \mathcal{H}_{sq_s} \tag{60}$$

so that the 1D projectors $P_{\mathcal{H}_{sq_s}} = \left| \xi_{sq_s} \right\rangle\left\langle \xi_{sq_s} \right|$ form the resolution of the projection operator $P_{\mathcal{H}_s}$ on subspace $\mathcal{H}_s$, i.e.,

$$P_{\mathcal{H}_s} = \sum_{q_s=1}^{L_s} P_{\mathcal{H}_{sq_s}} \tag{61}$$

Here, $L_s$ is the dimension of the $s^{th}$ subspace $\mathcal{H}_s$. Furthermore, the overall system Hilbert space $\mathcal{H}$ can be written as

$$\mathcal{H} = \bigoplus_{s=1}^{C} \bigoplus_{q_s=1}^{L_s} \mathcal{H}_{sq_s} \tag{62}$$

thus, identifying a resolution of the identity operator on $\mathcal{H}$ (more refined than that given by Eq. (20)) expressed as

$$I = \sum_{s=1}^{C} \sum_{q_s=1}^{L_s} P_{\mathcal{H}_{sq_s}} \tag{63}$$

Moreover, we have the orthogonality condition as

$$P_{\mathcal{H}_{sq_s}} P_{\mathcal{H}_{zq_z}} = \delta_{sz} \delta_{q_s q_z} P_{\mathcal{H}_{sq_s}} \tag{64}$$

Finally, the system-level Hamiltonian for $\mathcal{H}$ and that for each subspace $\mathcal{H}_s$ are given by

$$H = \sum_{s=1}^{C} H_s = \sum_{s=1}^{C} \sum_{q_s=1}^{L_s} E_{sq_s} P_{\mathcal{H}_{sq_s}} \tag{65}$$

*2.5 Occupation Number Operators*

The occupation number operators and the Hamiltonian operator can now be written so that the connection of the system-level operators to the one-particle operators can be made. As a first step, $N_{A_i j_i}^{s,int}$, which is the $A_i$-particles-in-the-$j_i^{th}$-internal-level occupation number operator on the internal part $\mathcal{H}_s^{int}$ of subspace $\mathcal{H}_s$, is expressed as

$$N_{A_i j_i}^{s, int} = \sum_{q_s^{int}=1}^{L_s^{int}} \alpha_{ij_i}^{sq_s^{int}} P_{\mathcal{H}_{sq_s^{int}}} \tag{66}$$

Multiplying this last expression by the identity operator on the translational part $\mathcal{H}_s^{tr}$ of subspace $\mathcal{H}_s$ yields

$$N_{A_i j_i}^s = N_{A_i j_i}^{s,int} \otimes I_s^{tr} = \sum_{q_s^{int}=1}^{L_s^{int}} \alpha_{ij_i}^{sq_s^{int}} P_{\mathcal{H}_{sq_s^{int}}} \otimes I_s^{tr} = \sum_{q_s^{tr}=1}^{L_s^{tr}} \sum_{q_s^{int}=1}^{L_s^{int}} \alpha_{ij_i}^{sq_s^{int}} P_{\mathcal{H}_{sq_s^{int}}} \otimes P_{\mathcal{H}_{sq_s^{tr}}} = \sum_{q_s=1}^{L_s} \alpha_{ij_i}^{sq_s} P_{\mathcal{H}_{sq_s}} \tag{67}$$

which is the $A_i$-particles-in-the-$j_i^{th}$-internal-level occupation number operator on the overall subspace $\mathcal{H}_s$ of the $s^{th}$ compatible composition. In the last equality, we have used the definition

$$\alpha_{ij_i}^{sq_s} = \alpha_{ij_i}^{sq_s^{int}} \text{ for every } s, i, j_i \tag{68}$$

which follows from the fact that since $q_s^{int}$ is the internal part of the full translational plus internal combinatorial labeling set $q_s = (q_s^{tr}, q_s^{int})$, specifying $q_s$ also specifies $q_s^{int}$, which in turn specifies the occupation numbers.

Therefore, at the overall system level, the $A_i$-particles-in-the-$j_i^{th}$-internal-level occupation number operator $N_{A_i j_i}$ for the overall Hilbert space $\mathcal{H}$, and the $A_i$-particles number operator $N_{A_i}$ are given by

$$N_{A_i j_i} = \sum_{s=1}^{C} N_{A_i j_i}^s \tag{69}$$

$$N_{A_i} = \sum_{j_i=1}^{M_i} N_{A_i j_i} = \sum_{s=1}^{C} \sum_{j_i=1}^{M_i} \sum_{q_s=1}^{L_s} \alpha_{ij_i}^{sq_s} P_{\mathcal{H}_{sq_s}} \tag{70}$$

Using Eqs. (10) and (61) in (70), the $A_i$-particles number operator $N_{A_i}$, which is not a c-number operator due to the fact that there are different eigenvalues $n_{is}$ for every one of the $C$ different compatible compositions, is written as

$$N_{A_i} = \sum_{s=1}^{C} n_{is} P_{\mathcal{H}_s} = \sum_{s=1}^{C} n_{is} \sum_{q_s=1}^{L_s} P_{\mathcal{H}_{sq_s}} \tag{71}$$

*2.6 Expectation Values of the Occupation Numbers, the Energy, and the Numbers of Particles*

The expectation value for the number of particles of species $A_i$ occupying the $j^{th}$-internal-one-particle eigenlevel can now be found from

$$\langle N_{A_i j_i} \rangle = \text{Tr}(\rho N_{A_i j_i}) \tag{72}$$

by defining the occupation probabilities of the system-level energy eigenlevels given by

$$y_{sq_s} = \text{Tr}(\rho P_{\mathcal{H}_{sq_s}}) = \langle \xi_{sq_s} | \rho | \xi_{sq_s} \rangle \tag{73}$$

Substitution of Eq. (67) and (69) into (72) and using (73) yields

$$\langle N_{A_i j_i} \rangle = \sum_{s=1}^{C} \sum_{q_s=1}^{L_s} y_{sq_s} \alpha_{ij_i}^{sq_s} \tag{74}$$

A similar result is obtained for the expectation value of the system energy $\langle H \rangle$, namely,

$$\langle H \rangle = \text{Tr}(\rho H) = \sum_{s=1}^{C} \sum_{q_s=1}^{L_s} y_{sq_s} E_{sq_s} \tag{75}$$

and for the expectation value for the number of particles of species $A_i$,

$$\langle N_{A_i} \rangle = \sum_{j_i=1}^{M_i} \langle N_{A_i j_i} \rangle = \sum_{s=1}^{C} \sum_{q_s=1}^{L_s} y_{sq_s} \sum_{j_i=1}^{M_i} \alpha_{ij_i}^{sq_s} = \sum_{s=1}^{C} \sum_{q_s=1}^{L_s} y_{sq_s} n_{is} \tag{76}$$

where we have used Eq. (10).

As a final note, the $\langle N_{A_i j_i} \rangle$ can be identified as the Bose-Einstein factors [43], $\alpha_{j_i}^{A_i}$, for each species $A_i$ and are valid not just at stable equilibrium but at each non-equilibrium state through which the system passes. Thus, they are more general then the closed-form expressions found in the literature [43-45]. It is also noteworthy that $y_{sq_s}$ represents the diagonal elements of the density matrix in the system-level Hamiltonian representation in which the full density matrix is, of course, given by

$$\langle \xi_{sq_s} | \rho | \xi_{zq_z} \rangle \tag{77}$$

and the off-diagonal elements are in general non-zero unless operators $\rho$ and $H$ commute.

*2.7 Position Representation of the Translational Hamiltonian and Its Eigenvectors*

Now, before proceeding to the dimensionalities and symmetries of the Hilbert spaces, we return to Eq. (52) and discuss how to compute the eigenvectors $\left| \hat{\varepsilon}_{k_{sj_s}}^{sj_s} \right\rangle$ of the transformed translational Hamiltonian $\hat{H}_{j_s}^s$. Here for simplicity of notation, we drop the $s$ and the $j_s$ subscripts and superscripts so that the eigenvectors are now $|\hat{\varepsilon}_k\rangle$ and the Hamiltonian is $\hat{H}$. We assume that the Hamiltonian for the reduced particle (or the center of mass in which case the potential energy function would be zero) has the usual form

$$\hat{H} = \frac{\hat{P}_X^2 + \hat{P}_Y^2 + \hat{P}_Z^2}{2\hat{m}} + V(\hat{X}, \hat{Y}, \hat{Z}) \tag{78}$$

where $\hat{m}$ is the reduced mass (or in the case of the center of mass, the total mass) and $\hat{P}_X^2, \hat{P}_Y^2$, and $\hat{P}_Z^2$ and $\hat{X}$, $\hat{Y}$, and $\hat{Z}$ are the momentum and position operators. Since the physical particles are confined in a 3D box of dimensions $a_x$, $a_y$, $a_z$, the reduced particles are also confined in a box of dimensions $a_x$, $a_y$, $a_z$. So the position operator $\underline{\hat{R}} = (\hat{X}, \hat{Y}, \hat{Z})$ has eigenvalue problem

$$\underline{\hat{R}} |\underline{\hat{r}}\rangle = \underline{\hat{r}} |\underline{\hat{r}}\rangle \tag{79}$$

where the eigenvalues are the possible positions $\underline{\hat{r}} = (\hat{x}, \hat{y}, \hat{z})$ in this box. The eigenvectors $|\underline{\hat{r}}\rangle = |\hat{x}, \hat{y}, \hat{z}\rangle$ form the standard convenient basis to represent vectors as functions in $\Re^3$. Such a position representation is easily constructed by using the resolution of the identity operator provided by the eigenvectors, i.e.,

$$\hat{I} = \iiint d\hat{x}\, d\hat{y}\, d\hat{z}\, |\underline{\hat{r}}\rangle\langle\underline{\hat{r}}| \tag{80}$$

together with the orthogonality condition

$$\langle \underline{\hat{r}} | \underline{\hat{r}}' \rangle = \delta(\underline{\hat{r}} - \underline{\hat{r}}') = \delta(\hat{x} - \hat{x}')\delta(\hat{y} - \hat{y}')\delta(\hat{z} - \hat{z}') \tag{81}$$

Thus, with respect to the position eigenbasis, the eigenvector $|\hat{\varepsilon}_k\rangle$ has "coordinates" $\hat{\varepsilon}_k(\underline{\hat{r}}) = \langle \underline{\hat{r}} | \hat{\varepsilon}_k \rangle$ in the sense that

$$|\hat{\varepsilon}_k\rangle = \iiint d\hat{x}\, d\hat{y}\, d\hat{z}\, \hat{\varepsilon}_k(\underline{\hat{r}}) |\underline{\hat{r}}\rangle \tag{82}$$

Similarly, the operator $\hat{H}$ has "matrix elements" $\hat{H}(\underline{\hat{r}}, \underline{\hat{r}}') = \langle \underline{\hat{r}} | \hat{H} | \underline{\hat{r}}' \rangle$ in the sense that

$$\hat{H} = \iiint d\hat{x}\, d\hat{y}\, d\hat{z} \iiint d\hat{x}'\, d\hat{y}'\, d\hat{z}'\, \hat{H}(\underline{\hat{r}}, \underline{\hat{r}}') |\underline{\hat{r}}\rangle\langle\underline{\hat{r}}'| \tag{83}$$

and so, for example, the vector $|\psi_k\rangle = \hat{H}|\hat{\varepsilon}_k\rangle$ has coordinates given by

$$\psi_k(\underline{\hat{r}}) = \iiint d\hat{x}'\, d\hat{y}'\, d\hat{z}'\, \hat{H}(\underline{\hat{r}}, \underline{\hat{r}}')\hat{\varepsilon}_k(\underline{\hat{r}}') \tag{84}$$

Notice that the orthogonality of the eigenvectors, i.e., the condition $\langle \hat{\varepsilon}_j | \hat{\varepsilon}_k \rangle = \delta_{jk}$, implies the orthogonality of the corresponding coordinate functions, namely,

$$\iiint d\hat{x} d\hat{y} d\hat{z} \hat{\varepsilon}_j^*(\hat{\underline{r}}) \hat{\varepsilon}_k(\hat{\underline{r}}) = \delta_{jk} \tag{85}$$

where the * indicates complex conjugate. Of course, the matrix elements of the position operators are $\underline{\hat{R}}(\hat{\underline{r}}, \hat{\underline{r}}') = \hat{\underline{r}} \delta(\hat{\underline{r}} - \hat{\underline{r}}')$, i.e., $\hat{X}(\hat{\underline{r}}, \hat{\underline{r}}') = \hat{x} \delta(\hat{\underline{r}} - \hat{\underline{r}}')$ and so on.

Due to the position-momentum commutation relations $[\hat{X}, \hat{P}_X] = [\hat{Y}, \hat{P}_Y] = [\hat{Z}, \hat{P}_Z] = i\hbar I$, the "matrix elements" of the momentum operators can be written as

$$\underline{\hat{P}}(\hat{\underline{r}}, \hat{\underline{r}}') = \delta(\hat{\underline{r}} - \hat{\underline{r}}') \frac{\hbar}{i} \frac{\partial}{\partial \hat{\underline{r}}} \tag{86}$$

i.e.,
$$\hat{P}_X(\hat{\underline{r}}, \hat{\underline{r}}') = \delta(\hat{\underline{r}} - \hat{\underline{r}}') \frac{\hbar}{i} \frac{\partial}{\partial \hat{x}} \tag{87}$$

$$\hat{P}_X^2(\hat{\underline{r}}, \hat{\underline{r}}') = \delta(\hat{\underline{r}} - \hat{\underline{r}}') \frac{\hbar}{-1} \frac{\partial^2}{\partial \hat{x}^2} \tag{88}$$

and so on so that we get the expression for the Hamiltonian matrix elements

$$\hat{H}(\hat{\underline{r}}, \hat{\underline{r}}') = \delta(\hat{\underline{r}} - \hat{\underline{r}}') \left[ -\frac{\hbar^2}{2\hat{m}} \hat{\nabla}_{\hat{\underline{r}}}^2 + V(\hat{\underline{r}}) \right] \tag{89}$$

Using the above relations, the eigenvalue problem $\hat{H}|\hat{\varepsilon}_k\rangle = e_k|\hat{\varepsilon}_k\rangle$ (i.e., Eq. (52)) is equivalent to the following:

$$-\frac{\hbar^2}{2\hat{m}} \hat{\nabla}_{\hat{\underline{r}}}^2 (\hat{\varepsilon}_k(\hat{\underline{r}})) + V(\hat{\underline{r}}) \hat{\varepsilon}_k(\hat{\underline{r}}) = e_k \hat{\varepsilon}_k(\hat{\underline{r}}) \tag{90}$$

Once this eigenvalue problem is solved for the eigenfunctions $\hat{\varepsilon}_k(\hat{\underline{r}})$, the orthogonality conditions (Eq. (85)) must be checked after which one can insert them into Eq. (82) to obtain the vectors $|\hat{\varepsilon}_k\rangle$.

*2.8 Dimensionalities and Symmetries of the Hilbert Spaces*

Before defining the dimensionalities of our various spaces, we make two assumptions. The first is that the particles behave as bosons (the extension to fermions is straightforward) and the second is that they are indistinguishable. The latter imposes that each factor space of identical and non-interacting degrees of freedom, i.e., each $\left(\mathcal{H}_{int}^{A_i}\right)^{\otimes n_{is}}$, be restricted to its symmetric subspace

$$\mathcal{B}_{int}^{n_{is} A_i} = \text{Symm}\left(\left(\mathcal{H}_{int}^{A_i}\right)^{\otimes n_{is}}\right) \tag{91}$$

where the symmetry is with respect to the interchange of two particles, which are, therefore, bosons. The subspace $\mathcal{B}_{int}^{n_{is} A_i}$ is in fact the eigenspace of the particle-exchange operation belonging to eigenvalue +1, meaning that applying such an exchange operation to any vector in $\mathcal{B}_{int}^{n_{is} A_i}$ yields the same vector multiplied by +1, i.e., leaves it unchanged. For fermions, we should instead be restricted to the eigenspace of the particle-exchange operation belonging to the eigenvalue −1,

$$\mathcal{F}_{int}^{n_{is} A_i} = \text{Anti-symm}\left(\left(\mathcal{H}_{int}^{A_i}\right)^{\otimes n_{is}}\right) \tag{92}$$

in which the particle-exchange operation changes the sign of any vector. Here, however, we deal only with bosons.

Therefore, each compatible composition subspace, Eq. (19), gets restricted to

$$\mathcal{H}_s^{\mathcal{B}} = \mathcal{H}_s^{tr} \otimes \mathcal{H}_s^{\mathcal{B} \, int} = \mathcal{H}_s^{tr} \bigotimes_{i=1}^{r} \mathcal{B}_{int}^{n_{is} A_i} \tag{93}$$

and the overall Hilbert space, Eq. (12), is restricted to

$$\mathcal{H}^{\mathcal{B}} = \bigoplus_{s=1}^{C} \mathcal{H}_s^{\mathcal{B}} = \bigoplus_{s=1}^{C} \left( \mathcal{H}_s^{tr} \otimes \bigotimes_{i=1}^{r} \mathcal{F}_{int}^{n_{is} A_i} \right) \tag{94}$$

Indeed, we recall that, for each $s$, the space $\mathcal{H}_s = \mathcal{H}_s^{tr} \otimes \mathcal{H}_s^{int}$ is spanned by the $s^{th}$ subset of eigenvectors $|\xi_{sq_s}\rangle$ of the Hamiltonian, which by our construction can be factored into the translational and internal eigenvectors as $|\xi_{sq_s}\rangle = |\xi_{sq_s}^{tr}\rangle \otimes |\xi_{sq_s}^{int}\rangle$. We assume that the translational factor space is already symmetrized by the assumption of pairwise particle interactions and the center-of-mass-reduced-masses procedure described above so that the dimension of $\mathcal{H}_s^{tr}$ is

$$L_s^{tr} = \prod_{j_s=1}^{n_s} K_{s j_s} \tag{95}$$

The internal factor space instead must be restricted, and this is effectively done by introducing the occupation numbers $\alpha_{i j_i}^{sq_s^{int}}$ and not counting repetitions so that while the dimension of $\mathcal{H}_s^{int}$ is

$$L_s^{int} = \prod_{i=1}^{r} M_i^{n_{is}} \tag{96}$$

that of $\mathcal{H}_s^{\mathcal{B} \, int} = \bigotimes_{i=1}^{r} \mathcal{F}_{int}^{n_{is} A_i}$ is

$$L_s^{\mathcal{B} \, int} = \prod_{i=1}^{r} L_{n_{is}}^{M_i} \tag{97}$$

where $\quad L_{n_{is}}^{M_i} = \dfrac{(n_{is} + M_i - 1)!}{n_{is}! (M_i - 1)!} \tag{98}$

Here $L_{n_{is}}^{M_i}$ is the number of possible ways that $n_{is}$ indistinguishable $A_i$ particles can be distributed on the $M_i$ internal energy eigenlevels.

The above is illustrated using the two-reaction-mechanisms system and the values for $n_{is}$ in Table 2 above. Recalling that $i = 1,...,5$ and $s = 1,...,3$, one finds that

$$L_{n_{1s}}^{M_1} = \left\{ \frac{(4 + M_1 - 1)!}{4!(M_1 - 1)!}, \frac{(3 + M_1 - 1)!}{3!(M_1 - 1)!}, \frac{(3 + M_1 - 1)!}{3!(M_1 - 1)!} \right\} \tag{99}$$

$$L_{n_{2s}}^{M_2} = \left\{ \frac{(1 + M_2 - 1)!}{1!(M_2 - 1)!}, 1, 1 \right\} \tag{100}$$

$$L_{n_{3s}}^{M_3} = \left\{ 1, \frac{(1 + M_3 - 1)!}{1!(M_3 - 1)!}, 1 \right\} \tag{101}$$

$$L_{n_{4s}}^{M_4} = \left\{ 1, 1, \frac{(1 + M_4 - 1)!}{1!(M_4 - 1)!} \right\} \tag{102}$$

$$L_{n_{5s}}^{M_5} = \left\{ 1, 1, \frac{(1 + M_5 - 1)!}{1!(M_5 - 1)!} \right\} \tag{103}$$

For simplicity of illustration, we assume $M_1=2$, $M_2=2$, $M_3=2$, $M_4=2$, and $M_5=2$, so that the dimensionalities of $\mathcal{H}^{\mathcal{B} \, int}$, $\mathcal{H}_s^{\mathcal{B} \, int}$, and $\mathcal{F}_{int}^{n_{is} A_i}$ are as given in Table 3.

**Table 3.** Occupation coefficients array for the two-reaction-mechanism system.

| | | s | 1 | 2 | 3 |
|---|---|---|---|---|---|
| $i$ | $M_i$ | $L^{M_i}_{n_{is}}$ | $L^{M_i}_{n_{i1}}$ | $L^{M_i}_{n_{i2}}$ | $L^{M_i}_{n_{i3}}$ |
| 1 | 2 | $L^{M_1}_{n_{1s}}$ | 5 | 4 | 4 |
| 2 | 2 | $L^{M_2}_{n_{2s}}$ | 2 | 1 | 1 |
| 3 | 2 | $L^{M_3}_{n_{3s}}$ | 1 | 2 | 1 |
| 4 | 2 | $L^{M_4}_{n_{4s}}$ | 1 | 1 | 2 |
| 5 | 2 | $L^{M_5}_{n_{5s}}$ | 1 | 1 | 2 |
| | | $L^{\mathscr{B}\ int}_s$ | 10 | 8 | 16 | $L=34$ |

Thus, for the two-reaction-mechanisms system, a 34×10 array of occupation coefficients $\alpha^{sq^{int}_s}_{ij_i}$ results made up of 3 smaller arrays, one for each subspace $s$, i.e., 10×10 for $\mathscr{H}^{\mathscr{B}\ int}_1$, 8×10 for $\mathscr{H}^{\mathscr{B}\ int}_2$, and 16×10 for $\mathscr{H}^{\mathscr{B}\ int}_3$. The arrays are shown in Table 4.

*2.9 Reaction Coordinate and System-Level Energy Occupation Probabilities*

Having defined a set of reaction coordinate occupation probabilities $w_s$ and a set $y_{sq_s}$ for the system-level energy and particle number eigenlevels, the relationship of the latter to the former must be established. Using Eqs. (34), (61), and (73), $w_s$ is written as

$$w_s = \text{Tr}(\rho P_{\mathscr{H}_s}) = \sum_{q_s=1}^{L_s} \text{Tr}(\rho P_{\mathscr{H}_{sq_s}}) = \sum_{q_s=1}^{L_s} y_{sq_s} \tag{104}$$

With this result and Eq. (34) and consistent with Eqs. (63) and (73), we have that

$$\sum_{s=1}^{C} w_s = \sum_{s=1}^{C} \sum_{q_s=1}^{L_s} y_{sq_s} = 1 \tag{105}$$

The expectation values for the reaction coordinate and reaction rate vectors can now be expressed in terms of the one-particle occupation probabilities using Eqs. (35), (36), and (104), i.e.,

$$\langle \mathbf{N} \rangle = \sum_{s=1}^{C} \mathbf{n}_s \sum_{q_s=1}^{L_s} y_{sq_s} \tag{106}$$

$$\langle \boldsymbol{\varepsilon} \rangle = \sum_{s=1}^{C} \boldsymbol{\varepsilon}_s \sum_{q_s=1}^{L_s} y_{sq_s} \tag{107}$$

and on a rate basis

$$\langle \dot{\mathbf{N}} \rangle = \sum_{s=1}^{C} \mathbf{n}_s \sum_{q_s=1}^{L_s} \dot{y}_{sq_s} \tag{108}$$

$$\langle \dot{\boldsymbol{\varepsilon}} \rangle = \sum_{s=1}^{C} \boldsymbol{\varepsilon}_s \sum_{q_s=1}^{L_s} \dot{y}_{sq_s} \tag{109}$$

**Table 4.** Occupation coefficients array for the two-reaction-mechanism system.

| | | | | $M_1=2$ | | $M_2=2$ | | $M_3=2$ | | $M_4=2$ | | $M_5=2$ | |
|---|---|---|---|---|---|---|---|---|---|---|---|---|---|
| | | | $j_i$ | 1 | 2 | 1 | 2 | 1 | 2 | 1 | 2 | 1 | 2 |
| | | $k$ | $q_s$ | | | | | | | | | | |
| $s=1$ $L_1^{\mathscr{B}\,int}=10$ | | 1 | 1 | 4 | 0 | 1 | 0 | 0 | 0 | 0 | 0 | 0 | 0 |
| | | 2 | 2 | 0 | 4 | 0 | 1 | 0 | 0 | 0 | 0 | 0 | 0 |
| | | 3 | 3 | 3 | 1 | 1 | 0 | 0 | 0 | 0 | 0 | 0 | 0 |
| | | 4 | 4 | 1 | 3 | 0 | 1 | 0 | 0 | 0 | 0 | 0 | 0 |
| | | 5 | 5 | 2 | 2 | 1 | 0 | 0 | 0 | 0 | 0 | 0 | 0 |
| | | 6 | 6 | 4 | 0 | 0 | 1 | 0 | 0 | 0 | 0 | 0 | 0 |
| | | 7 | 7 | 0 | 4 | 1 | 0 | 0 | 0 | 0 | 0 | 0 | 0 |
| | | 8 | 8 | 3 | 1 | 0 | 1 | 0 | 0 | 0 | 0 | 0 | 0 |
| | | 9 | 9 | 1 | 3 | 1 | 0 | 0 | 0 | 0 | 0 | 0 | 0 |
| | | 10 | 10 | 2 | 2 | 0 | 1 | 0 | 0 | 0 | 0 | 0 | 0 |
| $s=2$ $L_2^{\mathscr{B}\,int}=8$ | | 11 | 1 | 3 | 0 | 0 | 0 | 1 | 0 | 0 | 0 | 0 | 0 |
| | | 12 | 2 | 0 | 3 | 0 | 0 | 0 | 1 | 0 | 0 | 0 | 0 |
| | | 13 | 3 | 2 | 1 | 0 | 0 | 1 | 0 | 0 | 0 | 0 | 0 |
| | | 14 | 4 | 1 | 2 | 0 | 0 | 0 | 1 | 0 | 0 | 0 | 0 |
| | | 15 | 5 | 3 | 0 | 0 | 0 | 1 | 0 | 0 | 0 | 0 | 0 |
| | | 16 | 6 | 0 | 3 | 0 | 0 | 0 | 1 | 0 | 0 | 0 | 0 |
| | | 17 | 7 | 2 | 1 | 0 | 0 | 0 | 1 | 0 | 0 | 0 | 0 |
| | | 18 | 8 | 1 | 2 | 0 | 0 | 1 | 0 | 0 | 0 | 0 | 0 |
| $s=3$ $L_3^{\mathscr{B}\,int}=16$ | | 19 | 1 | 3 | 0 | 0 | 0 | 0 | 0 | 1 | 0 | 1 | 0 |
| | | 20 | 2 | 0 | 3 | 0 | 0 | 0 | 0 | 0 | 1 | 0 | 1 |
| | | 21 | 3 | 2 | 1 | 0 | 0 | 0 | 0 | 1 | 0 | 1 | 0 |
| | | 22 | 4 | 1 | 2 | 0 | 0 | 0 | 0 | 0 | 1 | 0 | 1 |
| | | 23 | 5 | 3 | 0 | 0 | 0 | 0 | 0 | 1 | 0 | 0 | 1 |
| | | 24 | 6 | 0 | 3 | 0 | 0 | 0 | 0 | 0 | 1 | 1 | 0 |
| | | 25 | 7 | 2 | 1 | 0 | 0 | 0 | 0 | 1 | 0 | 0 | 1 |
| | | 26 | 8 | 1 | 2 | 0 | 0 | 0 | 0 | 0 | 1 | 1 | 0 |
| | | 27 | 9 | 3 | 0 | 0 | 0 | 0 | 0 | 0 | 1 | 0 | 1 |
| | | 28 | 10 | 0 | 3 | 0 | 0 | 0 | 0 | 1 | 0 | 1 | 0 |
| | | 29 | 11 | 2 | 1 | 0 | 0 | 0 | 0 | 0 | 1 | 0 | 1 |
| | | 30 | 12 | 1 | 2 | 0 | 0 | 0 | 0 | 1 | 0 | 1 | 0 |
| | | 31 | 13 | 3 | 0 | 0 | 0 | 0 | 0 | 0 | 1 | 1 | 0 |
| | | 32 | 14 | 0 | 3 | 0 | 0 | 0 | 0 | 1 | 0 | 0 | 1 |
| | | 33 | 15 | 2 | 1 | 0 | 0 | 0 | 0 | 0 | 1 | 1 | 0 |
| | | 34 | 16 | 1 | 2 | 0 | 0 | 0 | 0 | 1 | 0 | 0 | 0 |

*2.10 One-particle Energy Eigenvalue Problems*

The previous development was necessitated by the need to relate the system-level energy eigenvalue problem (Eq. (2)) to a set of one-particle energy eigenvalue problems that can be solved more easily since the computational difficulties of solving the former, which represents a multi-body problem, quickly augment as the number of particles in the system increases. In fact, it quickly may become impossible to solve. Similar one-particle approaches have been used in statistical thermodynamics (ST) [44, 45] to derive stable equilibrium property relations for ideal, perfect, and Sommerfeld gases. When augmented by a set of multi-body, inter-particle potentials, this same approach leads to, for example, stable equilibrium property relations for dense gases (e.g., the virial equation of state [43-45]), liquids, and even solids. In SEAQT, such an approach has also been used to model the non-equilibrium time evolution of the state of a non-reacting system of a few hydrogen molecules flowing into a carbon nano-tube [46].

The general form of the one-particle energy eigenvalue problem for each species $A_i$ is given in Eq. (47) above for each of the internal modes (e.g., rotation, vibration, vibration-rotation, electronic, etc.) and by Eq. (52) for the translational mode. Separating the internal modes such as rotation and vibration is a good approximation when the amplitudes of vibration of the atoms of a molecule are small compared with the equilibrium distances between the atoms. This would, for example, be the case for the lower energy eigenstates. It also requires that the forces between atoms induced by the rotation are small when compared with the interatomic forces giving rise to the vibrations [43-45, 47]. In general such a separation works for so-called rigid molecules (e.g., water, carbon dioxide, methane, ethylene, benzene, etc.) but must be considered on a case-by-case basis for so-called non-rigid molecules (e.g., propane, ammonia, ethanol, water dimer, etc.). In the case of the latter, rotational-vibrational coupling may need to be taken into account.

Based on our development in *Section* 2.7 above, the translational eigenvalue problem (Eq. (52)) can be written in terms of eigenfunctions as is done in Eq. (90) and as is repeated here in slightly different form, i.e.,

$$\left(-\frac{\hbar^2}{2\hat{m}}\hat{\nabla}_{\hat{r}}^2 + V(\hat{r})\right)\hat{\varepsilon}_k(\hat{r}) = e_k\hat{\varepsilon}_k(\hat{r}) \tag{110}$$

where as before for simplicity of notation, we drop the $s$ and the $j_s$ subscripts and superscripts. In a similar fashion, Eq. (47) for the internal modes is expressed in terms of eigenfunctions as

$$\left(-\frac{\hbar^2}{2m}\nabla_r^2 + V(r)\right)\varepsilon_j(r) = e_j\varepsilon_j(r) \tag{111}$$

where again the subscript and superscript notation has been simplified. Note that the vector $r$ here is expressed by the set of coordinates most convenient for a particular internal mode. Thus, for example, the energy eigenvalue problem for the vibration of a diatomic atom in a single direction can be written in terms of a 1D harmonic oscillator so that the vector $r$ reduces to the single coordinate $x$. For any polyatomic molecule comprised of $n$ atoms, there are either $3n$-5 or $3n$-6 vibrational modes of freedom (MOF), i.e., a reduction of the $3n$ vibrational MOF to account for the three translational MOF, which describe the motion of the molecule's center of mass, plus either two or three rotational MOF, depending on whether or not the atoms are aligned. Thus, the vibrational eigenvalue problem may be written as the sum of $3n$-5 or $3n$-6 1D harmonic oscillator problems, the solution of which results in the following expression for the one-particle vibrational energy eigenvalue of the polyatomic molecule

$$e_{j_{vib}} = \sum_{j=1}^{3n-a}(j+1/2)h\nu_j \tag{112}$$

Here $a$ is either 5 or 6, $h$ is Planck's constant, and $\nu_j$ is the frequency of the $j^{th}$ vibrational mode.

In a similar fashion, the rotational energy eigenvalue problem for a polyatomic molecule can be written for a rigid polyatomic molecule in terms of the rigid rotor (rotating top) problem for which the vector $r$ is given in terms of the rotational angles in three directions, i.e., $r = (\theta, \varphi, \zeta)$. The rotational

Hamiltonian can then be expressed in a variety of coordinate frames but always depends upon the moments of inertia of the molecule about its center of gravity. By choosing the coordinates to lie along the principal axes of inertia of the body, the Hamiltonian depends only upon the principal moments of inertia $I_A$, $I_B$, and $I_C$. If then the rigid rotor representing the molecule is that of an oblate ($I_A = I_B < I_C$) symmetric rotating top, the solution of the rotational eigenvalue problem results in the following expression for the one-particle rotational energy eigenvalue of the polyatomic molecule:

$$e_{j_{rot}} = \frac{j(j+1)\hbar^2}{2I_A} + \left(\frac{1}{I_C} - \frac{1}{I_A}\right)\frac{l^2\hbar^2}{2} \qquad (113)$$

Here $\hbar$ is Planck's modified constant and $j$ a positive integer that must be at least as large as the absolute values of the quantum numbers $l$ and $m$, whichever of these is greater. Both $l$ and $m$ may have a range of positive and negative integer values. For a spherical symmetric top, such as would be used to represent a monoatomic molecule, all three moments of inertia are equal; and the second term on the right hand side of Eq. (113) vanishes. For a prolate symmetric top ($I_A < I_B = I_C$), $I_C$ replaces $I_A$ in Eq. (113) and $I_A$ replaces $I_C$. Finally for the case of asymmetric tops ($I_A \neq I_B \neq I_C$), Eq. (113) no longer holds and finding the rotational energy eigenvalue as a function of the moments of inertia is a much more involved process [47].

Finally, in the case of translation, the one-particle energy eigenvalue problem, Eq. (110), is that of the particle in a box. The solution of this problem results in the following expression for the one-particle translational energy eigenvalues

$$e_k = \frac{h^2}{8m}\left(\left(\frac{n_x}{L_x}\right)^2 + \left(\frac{n_y}{L_y}\right)^2 + \left(\frac{n_z}{L_z}\right)^2\right) \qquad (114)$$

where $k=1,2,\ldots$ is the principal quantum number; $n_x$, $n_y$ and $n_z$ are the quantum numbers in the $x$, $y$ and $z$ directions, respectively; and $L_x$, $L_y$, and $L_z$ are the dimensions of the system volume in the $x$, $y$, and $z$ directions, respectively. It is this expression and the previous two which are used here to represent the one-particle energy eigenstructure of the constituents of the Gibbs-Dalton mixture of ideal gases, which we have assumed for the reacting mixture constrained by the two reaction mechanisms (13) and (14). We next consider the system of equations formed by the equation of motion of SEAQT, which is used to predict the chemical kinetics of this two-reaction-mechanism system.

## 3. SEAQT dynamical model equation

The SEAQT equation of motion for the density operator of a single assembly of indistinguishable particles system can be expressed as follows [48]:

$$\frac{d\rho}{dt} = -\frac{i}{\hbar}[H,\rho] + \frac{1}{2k_B\tau}\{\Delta M, \rho\} \qquad (115)$$

where $\tau$ is a time constant[1] and $\{\Delta M, \rho\}/2k_B\tau$ the so-called dissipation term. The latter is defined precisely below. It is a function of $\rho$, $\ln\rho$, and $H$ designed so as to capture the nonlinear dynamics of an irreversible process by pulling the density or state operator $\rho$ in the direction of the projection of the gradient of the von Neumann entropy functional, i.e.,

$$\langle S \rangle = -k_B \text{Tr}(\rho \ln \rho) \qquad (116)$$

---

[1] As discussed in [1,13,46], when the SEA geometrical construction is done with respect to the Fisher-Rao metric, which is the version we adopt here for simplicity, the parameter $\tau$ can in general be a positive functional of the state $\rho$. However, here, again for simplicity, we assume it to be a constant.

onto the manifold of constant (expectation value of the system's) energy. Note that the conservation of atomic elements is already built into the structure of the Hilbert space as a direct sum of subspaces each belonging to a fixed composition stoichiometrically compatible with the initial composition via the proportionality relations, Eq. (5). The dissipation operator in Eq. (115) is written as

$$\{\Delta M, \rho\} = \Delta M \rho + \rho \Delta M \tag{117}$$

where $M$ denotes the nonequilibrium Massieu operator [13, 22, 48] defined by

$$M = S - H/\theta_H(\rho) \tag{118}$$

$$\theta_H(\rho) = \langle \Delta H \Delta H \rangle / \langle \Delta S \Delta H \rangle \tag{119}$$

$\Delta H$ and $\Delta S$ are the deviation operators of $H$ and $S$ given by

$$\Delta H = H - I\langle H \rangle \tag{120}$$

$$\Delta S = S - I\langle S \rangle \tag{121}$$

and the $S$ operator is expressed as either of the two equivalent forms

$$S = -k_B \ln(\rho + P_o) = -k_B B \ln \rho \tag{122}$$

with $P_o$ and $B$ are, respectively, the projection operators onto the kernel and the range of $\rho$.

Eq. (115) has been proven to be compatible with both the first and the second laws of thermodynamics [8-10]. In particular, the rate of entropy generation is a positive, semi-definite (nonlinear) functional of $\rho$ given by the following equivalent expressions:

$$\frac{d\langle S \rangle}{dt} = \frac{1}{k_B \tau} \langle \Delta M \Delta M \rangle = \frac{1}{k_B \tau} \left( \langle \Delta S \Delta S \rangle - \frac{\langle \Delta H \Delta H \rangle}{\theta_H} \right) \tag{123}$$

where

$$\langle \Delta H \Delta H \rangle = \text{Tr}(\rho \Delta H^2) = \text{Tr}(\rho H^2) - (\text{Tr}(\rho H))^2 \tag{124}$$

$$\langle \Delta S \Delta S \rangle = \text{Tr}(\rho \Delta S^2) = \text{Tr}(\rho S^2) - (\text{Tr}(\rho S))^2 \tag{125}$$

and

$$\langle \Delta S \Delta H \rangle = \text{Tr}(\rho \Delta S \Delta H) = \text{Tr}(\rho S H) - \text{Tr}(\rho S)\text{Tr}(\rho H) \tag{126}$$

Now, we reformulate the equation of motion in terms of the occupation probabilities $y_{sq_s}$ and the system-level energy eigenvalues $E_{sq_s}$ and the eigenprojectors $P_{\mathcal{H}_{sq_s}}$. Using Eqs. (63), (65), and (73), the first term on the right of Eq. (124) can be written as

$$\text{Tr}(\rho H^2) = \sum_{s=1}^{C} \sum_{q_s=1}^{L_s} y_{sq_s} (E_{sq_s})^2 \tag{127}$$

while the second term results in

$$(\text{Tr}(\rho H))^2 = \left( \sum_{s=1}^{C} \sum_{q_s=1}^{L_s} y_{sq_s} E_{sq_s} \right)^2 \tag{128}$$

Thus, $\quad \langle \Delta H \Delta H \rangle = \sum_{s=1}^{C} \sum_{q_s=1}^{L_s} y_{sq_s} (E_{sq_s})^2 - \left( \sum_{s=1}^{C} \sum_{q_s=1}^{L_s} y_{sq_s} E_{sq_s} \right)^2 \tag{129}$

A similar development for $\langle \Delta S \Delta H \rangle$ yields

$$\langle \Delta S \Delta H \rangle = \sum_{s=1}^{C} \sum_{q_s=1}^{L_s} y_{sq_s} S_{sq_s} E_{sq_s} - \left( \sum_{s=1}^{C} \sum_{q_s=1}^{L_s} y_{sq_s} S_{sq_s} \right) \left( \sum_{s=1}^{C} \sum_{q_s=1}^{L_s} y_{sq_s} E_{sq_s} \right) \tag{130}$$

where we define

$$S_{sq_s} = -k_B \frac{\text{Tr}(P_{\mathcal{H}_{sq_s}} \rho \ln \rho)}{y_{sq_s}} \tag{131}$$

An important simplification obtains when the state operator $\rho$ commutes with $H$. Then, we have that

$$\rho = \sum_{s=1}^{C} \sum_{q_s=1}^{L_s} y_{sq_s} P_{\mathcal{H}_{sq_s}} \tag{132}$$

and $\quad S_{sq_s} = -k_B \ln(y_{sq_s}) \tag{133}$

so that Eq. (130) reduces to

$$\langle \Delta S \Delta H \rangle = \sum_{s=1}^{C} \sum_{q_s=1}^{L_s} y_{sq_s} E_{sq_s} (-k_B \ln y_{sq_s}) - \left( \sum_{s=1}^{C} \sum_{q_s=1}^{L_s} y_{sq_s} E_{sq_s} \right) \left( -k_B \sum_{s=1}^{C} \sum_{q_s=1}^{L_s} y_{sq_s} \ln y_{sq_s} \right) \tag{134}$$

Likewise, when $\rho$ and $H$ commute, Eq. (125) can be rewritten as

$$\langle \Delta S \Delta S \rangle = \sum_{s=1}^{C} \sum_{q_s=1}^{L_s} y_{sq_s} (-k_B \ln y_{sq_s})^2 - \left( -k_B \sum_{s=1}^{C} \sum_{q_s=1}^{L_s} y_{sq_s} \ln y_{sq_s} \right)^2 \tag{135}$$

We now return to the equation of motion and assume for simplicity that $\rho$ and $H$ commute (it can be proven [9] that if they commute at one instant of time, they commute at all times). Then, by multiplying both sides of Eq. (115) by $P_{\mathcal{H}_{sq_s}}$, taking the trace, and making use of the foregoing equations, we finally obtain the following SEAQT "master" equation for the occupation probabilities

$$2 k_B \tau \dot{y}_{sq_s} = 2 y_{sq_s} \left( -k_B \ln y_{sq_s} + k_B \sum_{s=1}^{C} \sum_{q_s=1}^{L_s} y_{sq_s} \ln y_{sq_s} - \frac{1}{\theta_H} \left( E_{sq_s} - \sum_{s=1}^{C} \sum_{q_s=1}^{L_s} y_{sq_s} E_{sq_s} \right) \right) \tag{136}$$

where $\quad \theta_H = \dfrac{\displaystyle\sum_{s=1}^{C} \sum_{q_s=1}^{L_s} y_{sq_s} (E_{sq_s})^2 - \left( \displaystyle\sum_{s=1}^{C} \sum_{q_s=1}^{L_s} y_{sq_s} E_{sq_s} \right)^2}{\displaystyle\sum_{s=1}^{C} \sum_{q_s=1}^{L_s} y_{sq_s} (E_{sq_s})(-k_B \ln y_{sq_s}) - \left( \displaystyle\sum_{s=1}^{C} \sum_{q_s=1}^{L_s} y_{sq_s} E_{sq_s} \right) \left( -k_B \displaystyle\sum_{s=1}^{C} \sum_{q_s=1}^{L_s} y_{sq_s} \ln y_{sq_s} \right)} \tag{137}$

Eq. (136) represents the SEAQT "master" equation providing the model dynamics when $[H,\rho]=0$. Once solved beginning from some initial state $\rho(0)$ with $[H,\rho(0)]=0$, it predicts the unique non-equilibrium thermodynamic path, which the reactive system follows in its relaxation towards the state of stable chemical equilibrium. An example of the application of this equation is given for a one-reaction mechanism system in [3, 37]. Its application to a complex set of coupled reaction pathways appears in [39, 40]. The implementation of Eq. (136) in this framework can in principle be used to predict the reaction rate constants when linked to experimental data found in the literature [49-51]. An application to the two-reaction mechanism system employed in this paper to explain the different elements of our modeling framework is provided by Al-Abbasi in the *Appendix* below as a means to illustrate the kind of results that can be obtained.

## 4. Conclusions

The principle of steepest entropy ascent in its quantum thermodynamics version provides an interesting modeling tool for the study of chemical kinetics at an atomistic-scale. The SEAQT framework starts from a fully quantum mechanical description of the properties of the participating species -- that must include reactants, products, and activated complexes as well -- taking into account the stoichiometry of an assumed set of governing reaction mechanisms between such species. It then represents the time evolution of the quantum state of the initial mixture as the solution of the nonlinear SEAQT master equation for the occupation probabilities of the overall system energy levels from which all other properties can be calculated as functions of time as the closed and isolated overall system evolves towards the final stable chemical equilibrium state. The extension of the method to model an open system is presented elsewhere (e.g., in [39, 40, 52]), but it is clear that for suitable steady boundary conditions the system will evolve towards a final steady state.

The modeling approach presented here holds the promise of providing at affordable computational costs a full set of thermodynamically consistent, time-dependent features of the chemical kinetics from an atomistic-scale point of view. Its extension to meso- and macro-scales has been made as well [3, 37, 39, 40, 52]. The formulation enjoys a built-in strong form of thermodynamic consistency, which is highly desirable in first principle modeling. It also prompts for generalizations in at least two important directions that seem to be worth further consideration. The first is to extend the method so as to include more time constants as will undoubtedly be necessary for modeling certain more complex chemical kinetic schemes. This can be done by choosing a more elaborate metric than the Fisher-Rao metric, as suggested in [1]. The second is to investigate in the chemical kinetic framework the effects of nonlocality and quantum decoherence, which can be done by choosing the composite-system SEAQT formulation along the lines already discussed in another quantum thermodynamic framework application [2] where the SEAQT principle has been shown to correctly incorporate the impossibility of producing and exploiting quantum entanglement by means of local or classical protocols.

**APPENDIX by O. Al-Abbasi**

*A1. Previous Work*

The reaction mechanisms, Eqs. (13) and (14), above are a decomposition of the single reaction mechanism,

$$F + H_2 \Leftrightarrow HF + H \tag{A-1}$$

the kinetic characteristics of which have been thoroughly studied both theoretically and experimentally in the literature for the last several decades. For example, Wilkins [53], Muckerman [54], and Hutchinson and Wyatt [55] have investigated the reaction rate constant of this reaction mechanism classically using the trajectory calculation approach. A number of quasi-classical studies have as well been conducted for this reaction (see, e.g., Feng, Grant, and Root [56] and Aoiz et al. [57, 58]).

This chemical reaction is considered as one of the few reactions that can be studied using extensive quantum models due to the simple geometrical configurations of the elemental constituents. The literature is rich with studies that investigate several aspects of the $F+H_2$ reaction from the quantum mechanical point of view. In particular, Wu, Johnson, and Levine [59] study the effects of the location of the energy barrier on the characteristics of the reaction dynamics. Redmon and Wyatt [60] and Rosenman et al. [61] predict the reaction probabilities and the cross-sections for the three-dimensional quantum model at a low region of electronic energies, while Wang, Thompson, and Miller [62] predict the rate constants on the highly accurate Stark-Werner potential energy surface using a time-dependent quantum mechanical approach. Castillo et al. [63] investigate the effect of collision angle on the cross-section of the reaction. A more recent study by Moix and Huarte-Larrañaga [64], investigates the rate constant based on flux correlation functions.

Among the many experimental studies of the $F + H_2$ reaction mechanism is the one conducted by Wurzberg and Houston [49] who study the reaction rate constant in the temperature range from 190 K to 373 K. Heidner et al. [50] investigate the reaction rate constant over a wider temperature range (i.e., from 295 K to 765 K), while Chapman et al. [65] look at the state-to-state reactive scattering process at collision energies of 2.4 kcal/mole. Other studies are devoted to estimating the activation energy (i.e., the height of the reaction barrier) of the chemical reaction [66-68]. For a review of this chemical reaction, one can refer to the work by Persky and Kornweitz [69].

*A2. Energy Eigenstructure*

The single reaction mechanism, Eq. (A-1), is decomposed in the two-reaction mechanism given by Eqs. (13) and (14) so as to include the effects of the activated complex *HHF* in the kinetic modeling. This model is used to demonstrate the SEAQT framework for a multi-reaction system developed in the bulk of the present paper.

The initial composition of the reacting mixture chosen for the results presented below is that of 1 particle of *F* and 1 of $H_2$. We assume an initial state of the mixture that in the absence of reactions would be a stable equilibrium state at 300 K. Note that this temperature plays no role in the kinetic predictions of the SEAQT framework other than that it is used to establish the initial state of the mixture. The relaxation time $\tau$ of the equation of motion (i.e., Eq. (136)) is kept constant and set equal to $5.2\times10^{-11}$ sec. This value results in predictions for the initial forward reaction rate constant comparable to those found in Al-Abbasi [38] as well as in the literature (e.g., Heidner et al. [50]).

The parameter values used to determine the energy eigenstructure of the reactive system are given in Table A1 for all constituents other than the activated complex *HHF*. For *HHF*, the wavenumbers employed are 397 cm$^{-1}$, 392 cm$^{-1}$, and 4007 cm$^{-1}$ [70]. The ranges of the polyatomic-constituent quantum numbers considered for the internal degrees of freedom are given in Table A2 and follow those of the energy eigenstructure considered in [71, 72]. The ranges of the translational quantum numbers for each of the constituents are given in Table A3. Following [38], the choice for these ranges has been chosen so that the energy eigenstructures for the reactants, activated complex, and products overlap.

**Table A1.** Parameter values assumed, following [38], to calculate the energy eigenstructure for the two-reaction mechanism system.

|  | *F* | $H_2$ | *H* | *HF* |
|---|---|---|---|---|
| **Mass** (kg) | 3.1548×10$^{-26}$ | 3.3475×10$^{-27}$ | 1.6737×10$^{-27}$ | 3.3221×10$^{-26}$ |
| **Bond length** ($\text{Å}$) | - | 0.7416 | - | 0.917 |
| **Dissociation energy** (eV) [73] | - | 4.48 | - | 5.85 |
| **Wavenumber** (cm$^{-1}$) [74] | - | 4401 | - | 4099 |

**Table A2.** Ranges of the internal energy eigenstructure quantum numbers assumed, following [38], for the polyatomic molecules of the two-reaction mechanism system.

|  | $H_2$ | *HF* | *HHF* |
|---|---|---|---|
| Vibrational quantum number | 0 | 0, 1, …, 3 | 0, 1, 2 |
| Rotational quantum number | 0, 1, 2, 3 | 0, 1, …, 10 | 0, 1 |

**Table A3.** Ranges of the translational energy eigenstructure assumed, following [38], for each species of the two-reaction mechanism system.

| $A_i$ | Translational quantum number (10 evenly spaced samples) |
|---|---|
| F | $k=1,...,5500$ |
| $H_2$ | $k=1,...,1000$ |
| HHF | $k=1,...,6500$ |
| HF | $k=1,...,5000$ |
| H | $k=1,...,1000$ |

*A3. SEAQT Results*

Figures A1 and A2 show how both the entropy and the rate of entropy generation of the system evolve in time. A state of stable chemical equilibrium is reached when the entropy plateaus out. In addition, the peak in the rate of entropy generation occurs quite early in the process and then quickly decreases as stable equilibrium is approached.

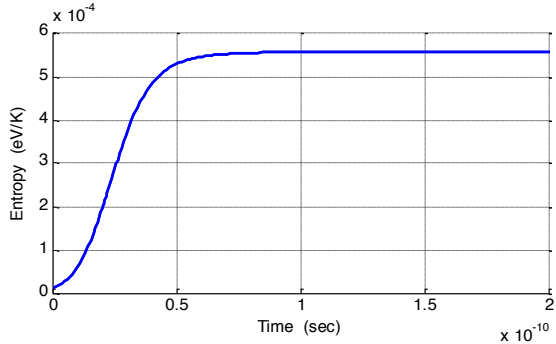 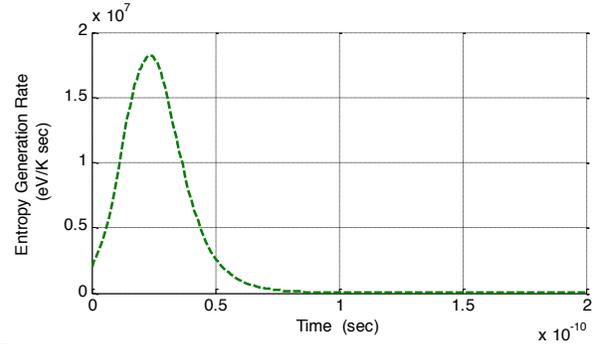

**Figure A1.** Expectation values of the entropy as a function of time obtained for the $F+H_2$ two-reaction mechanism system corresponding to an initial temperature of 300 K for the assumed energy eigenstructure.

**Figure A2.** Expectation values of the entropy generation rate as a function in time for the $F+H_2$ two-reaction mechanism system corresponding to an initial temperature of 300 K for the assumed energy eigenstructure.

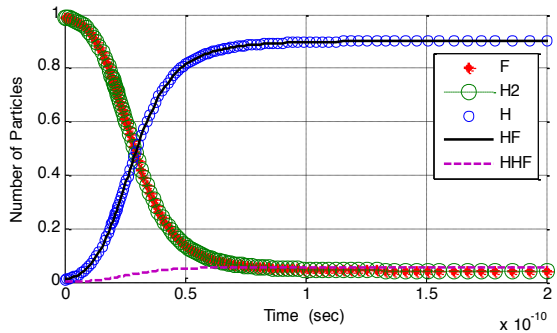 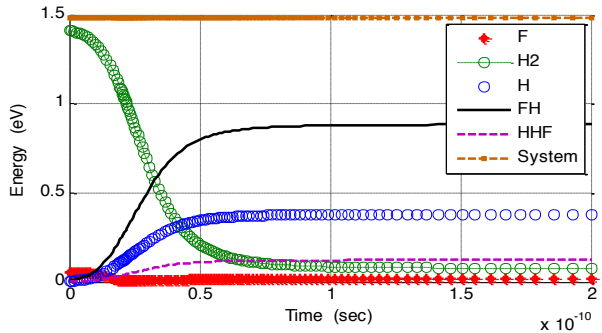

**Figure A3.** Expectation values of the particle number operator for each species for the $F+H_2$ two-reaction mechanism system corresponding to an initial temperature of 300 K for the assumed energy eigenstructure.

**Figure A4.** Expectation energies for each species and the overall system for the $F+H_2$ two-reaction mechanism system corresponding to an initial temperature of 300 K for the assumed energy eigenstructure.

It must be noted that the results presented here are of preliminary qualitative significance only, because of the rather limited number of internal and translational energy levels that we have considered (on the order of $10^3$) here to describe the energy eigenstructure of the reactive system. This can be corrected by using a more realistic energy eigenstructure, which, nevertheless, would consist of an extremely large number of eigenlevels (on the order of $10^{130}$). To this end, the use of the density of states approach proposed in [3, 24, 37] has shown to hold great promise for providing accurate quantitative results with little computational effort. As expected, a change in the number of energy eigenlevels also changes in turn the duration of the reaction process.

It is also important to point out that as noted in [1, 48] and further emphasized in [3], the kinetic SEA path of the chemical reaction process predicted by the SEAQT geometrical construction presented here is independent of the single relaxation time assumed in the model. No matter which relaxation time is used, the non-equilibrium path in state space remains the same. This result can be shown from the equation of motion where the single relaxation time does not influence the ratio of each pair of occupation probabilities. The relaxation time here has been fitted to match the experimental data in order to show the state-space path result in the proper time scale. It must also be noted that the SEAQT scheme can produce more elaborate models with multiple relaxation times by using a generalized form of the equation of motion, as discussed in [1], which from the geometrical point of view selects the SEA path in state space not with respect to the uniform Fisher-Rao metric as assumed here, but with respect to a more general metric tensor whose different principal values are related to the different relaxation times.

A key feature of the SEAQT framework is that it is able to dynamically predict the full time dependence of the concentrations of the various species as the reaction evolves in time. Figure A3 shows how the reactant species are depleted and the product species created along the entire reaction process. Indeed, the availability of these instantaneous values for the species concentrations indicates that the reaction rate constant $k(T_{\text{initial}})$, which in the literature is usually referred to as the "thermal rate constant" (i.e., the forward reaction rate of the overall reaction evaluated at the start of the process when the backward reaction rate is still negligible [75]), may, in fact, not be a constant but rather a parameter changing in time. In Figure A3, the identical amounts for the reactants $F$ and $H_2$ as well as for the products $H$ and $FH$ are direct consequences of the proportionality relations, Eq. (5), and of the particular initial amounts chosen for reactants and products.

Figure A4 shows the expectation values of the overall energy of the system (constant, due to energy conservation) and of the partial energies of the various species contributing to overall energy during the reaction process. The partial energies of the product species are the same or almost the same as those for the one-reaction-mechanism system. In contrast, the partial energies of the reactant species are somewhat smaller, compensating for the energy of the activated complex $HHF$.

For the first reaction mechanism, Eq. (13), the net reaction rate as a function of time $t$ is given by

$$r_1(t) = r_{f_1}(t) - r_{b_1}(t) = k_{f_1}(t)[F(t)][H_2(t)] - k_{b_1}(t)[HHF(t)] \tag{A-2}$$

while that for second, Eq. (14), it is given by

$$r_2(t) = r_{f_2}(t) - r_{b_2}(t) = k_{f_2}(t)[HHF(t)] - k_{b_2}(t)[HF(t)][H(t)] \tag{A-3}$$

Here, $r_f$ and $r_b$ are the forward and backward reaction rates, $k_f$ and $k_b$ the forward and backward reaction rate "constants", and the symbols $[A(t)]$ denote the time dependent concentrations of the various species. The reaction orders for the five species $F$, $H_2$, $HHF$, $FH$, and $H$ coincide in this instance with the respective stoichiometric coefficients (we mention this point because for more general reaction schemes this may not be the case [75]). Based on the chosen initial amounts and the proportionality relations, it follows that the vector $\{r_1, r_2\}$ formed by the two net reaction rates for the two-reaction mechanism system coincides with the expectation value $\langle \dot{\varepsilon} \rangle$ of the rate of change of the reaction coordinate vector, as given by Eq. (109) (recall also Eq. (25)). Numerically, it can also be found by calculating (for example, using a second order accurate finite difference scheme) the rates of change of the expectation values of the particle number operators of two independent species. Once the instantaneous values of $\{r_1(t), r_2(t)\}$ are obtained from the solution of the SEAQT equation of motion, they can be used to determine $k_f(t)$ and $k_b(t)$ at every instant of time along the entire kinetic path as follows. We use Eqs. (A-2) and (A-3) along with the zero rate condition at the final chemical equilibrium state for both mechanisms and the assumption that the detailed balance condition holds also for the time-dependent rate constants, based on the equilibrium constants evaluated at the calculated temperature $T_{se}$ of the final chemical equilibrium state, i.e., assuming

$$\frac{k_{f_1}(t)}{k_{b_1}(t)} = \frac{k_{f_1}(t_{se})}{k_{b_1}(t_{se})} = \frac{[F]_{se}[H_2]_{se}}{[HHF]_{se}} = K_1(T_{se}) \tag{A-4}$$

$$\frac{k_{f_2}(t)}{k_{b_2}(t)} = \frac{k_{f_2}(t_{se})}{k_{b_2}(t_{se})} = \frac{[HHF]_{se}}{[HF]_{se}[H]_{se}} = K_2(T_{se}) \tag{A-5}$$

Figures A5 and A6 show, respectively, these instantaneous values for the first and the second reaction mechanism as well as the equilibrium constants $K_1(T_{se})$ and $K_2(T_{se})$ used to specify the ratios of $k_f$ to $k_b$ according to Eqs. (A-4) and (A-5), respectively, for our initial composition of one particle of $F$ and one of $H_2$ initially at 300 K. As shown in Figure A5, for the first reaction the backward reaction rate constant is orders of magnitude higher than that for the forward reaction rate constant. The reason is that the concentration of the activated complex $HHF$ is much smaller compared to the concentration of reactants. Therefore, in order to satisfy the net rate, the backward rate constant $k_b$ must maintain a very high value. In contrast, for the second reaction, Figure A6 shows that the forward reaction rate constant has much higher values than the backward rate constant. The reason is that for the second reaction, it is $k_f$, which is associated with the concentration of the activated complex. Note that the values of $k_f$ and $k_b$ for the second reaction exhibit smaller variations compared to the first reaction, which is explained below in light of the time traces of the reaction rates.

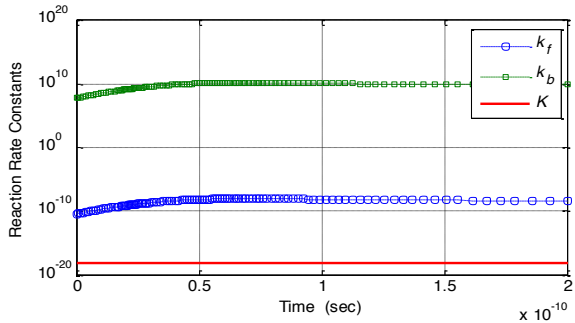 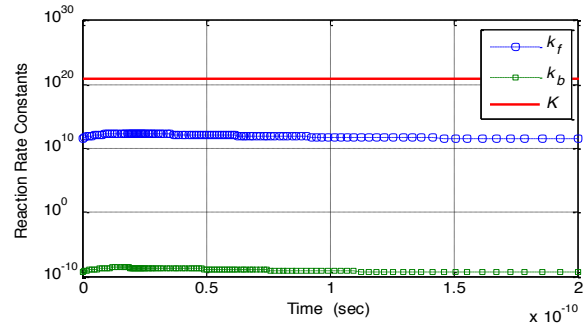

**Figure A5.** Forward and backward reaction rate constants $k_{f_1}(t)$ and $k_{b_1}(t)$ as well as the equilibrium constant $K_1(T_{se})$ as functions of time for the $F + H_2 \rightarrow HHF$ reaction mechanism for an initial composition of one particle of $F$ and one of $H_2$ at 300 K.

**Figure A6.** Forward and backward reaction rate constants $k_{f_2}(t)$ and $k_{b_2}(t)$ as well as the equilibrium constant $K_2(T_{se})$ as a function of time for the $HHF \rightarrow HF + H$ reaction mechanism for an initial composition of one particle of $F$ and one of $H_2$ at 300 K.

Figure A7 and A8 present the forward, backward, and net reaction rates for the two reactions, respectively. For the first reaction, Figure A7 shows that initially the forward reaction rate dominates and peaks at the same time as the entropy generation rate. In contrast, the backward reaction rate only becomes noticeable when the forward reaction rate has exhausted its peak. In contrast, for the second reaction, Figure A8 shows a markedly different behaviour. The forward and backward reaction rates are both relatively close in values even early in the process and continue to be so throughout the reaction. This explains why the reaction rate constants in Figure A6 are more flattened when compared to their values in Figure A5 for the first reaction mechanism.

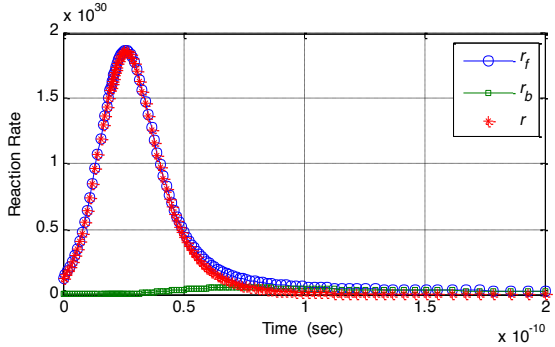 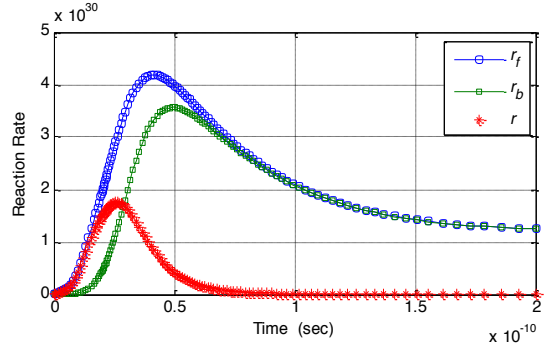

**Figure A7.** Forward, backward, and net reaction rates, $r_{f_1}(t)$, $r_{b_1}(t)$, and $r_1(t) = r_{f_1}(t) - r_{b_1}(t)$, as functions of time for the $F + H_2 \to HHF$ reaction mechanism for an initial composition of one particle of $F$ and one of $H_2$ at 300 K.

**Figure A8.** Forward, backward, and net reaction rates, $r_{f_2}(t)$, $r_{b_2}(t)$, and $r_2(t) = r_{f_2}(t) - r_{b_2}(t)$, as a function of time for the $HHF \to HF + H$ reaction mechanism for an initial composition of one particle of $F$ and one of $H_2$ at 300 K.

The difference between the energy of the activated complex at any given instant of time and that of the reactants, defined here as an activation energy, i.e.,

$$\langle E_A \rangle = \langle H_{HHF} \rangle - \langle H_F \rangle - \langle H_{H_2} \rangle \tag{A-6}$$

is reported in Figure A9 for three different initial temperatures: 300 K, 500 K, and 700 K. The first thing to note is that this activation energy is a dynamic property as opposed to the static activation energy usually calculated from information about the potential energy surface alone. Second, the height of the energy barrier in this analysis is an increasing function of the temperature of the initial mixture. In fact, Figure A9 shows that at around $2 \times 10^{-10}$ s the activation energy approaches a constant value equal to 29.9, 31.3 and 32.8 meV for an initial temperature of 300, 500, and 700 K, respectively. The time dependence is related to the amount of $HHF$ present at any given instant of time. Initially, when little $HHF$ is present, the value of the activation energy simply reflects the negative of the sum of the reactant energies present, which places the curve for 300 K above that for 500 K and that for 500 K above that for 700 K. Between $0.22 \times 10^{-10}$ and $0.241 \times 10^{-10}$ s this picture begins to change as the three curves cross over each other due to the appearance of a sufficient amount of the activated complex $HHF$, which for the higher temperatures is more significant. The amount of $HHF$ for the three cases continues to increase driving the activation energy towards positive values, which culminate in the values listed above for the final stable chemical equilibrium states.

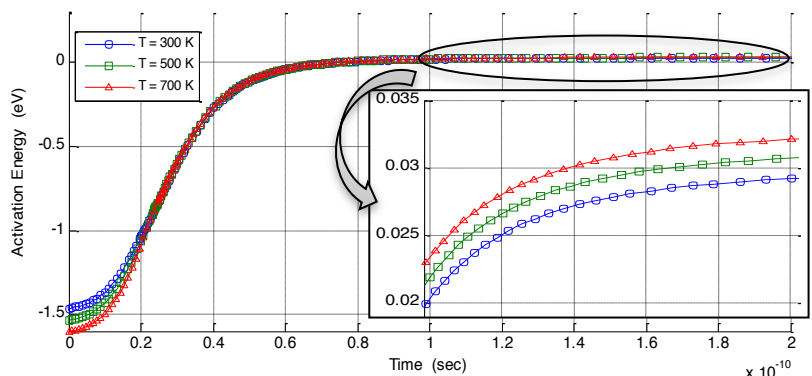

**Figure A9.** Time evolution of the activation energy defined by $\langle E_A \rangle = \langle H_{HHF} \rangle - \langle H_F \rangle - \langle H_{H_2} \rangle$ for three different initial temperatures of the initial 'mixture' of one particle of $F$ and one of $H_2$.

One final point to make here is in regard to the computational burden involved in these calculations. To date, the model predictions produced have been made on an Intel Duo Core CPU with 2.13 GHz workstation and are completed in a matter of seconds to minutes to a few hours for the reactions considered to date. These include not only the ones presented in this paper but much more complicated multi-reaction pathways such as those, for example, for the competing electrochemical reactions of oxygen and chromium oxide at the cathode electrode of a solid oxide fuel cell [39, 40]. Even in these cases, the computational burden is minimal. This represents an important advantage compared to the computational cost required by, for example, scattering calculations [76]. Furthermore, since the solution is obtained via solving a set of first order ordinary differential equations, the memory requirement for solving the system models formulated using this framework is minimal, which contrasts with methods that depend on 3D grids of the configurational space where the dimensionality grows exponentially [77]. For this reason and the others demonstrated above, the SEAQT framework appears to offer a great potential for contributing to interesting advances in the field of reaction kinetics.